\documentclass[12pt,leqno]{article}
\usepackage{amssymb,latexsym}
\usepackage{mathrsfs}
\usepackage[pdftex]{graphicx}
\usepackage[left=1in,right=1in,top=1in, bottom=1in]{geometry}
\usepackage{chicago}
\usepackage{amsmath}
\usepackage{graphicx}

\usepackage{caption} 
\usepackage{enumerate}
\usepackage{subcaption}
\usepackage{setspace}
 \usepackage{xcolor}
  \usepackage{booktabs,caption}
 \usepackage[flushleft]{threeparttable}

 \usepackage{xr}

\usepackage[hidelinks]{hyperref}

\long\def\symbolfootnote[#1]#2{\begingroup%
\def\thefootnote{\fnsymbol{footnote}}\footnote[#1]{#2}\endgroup}

\newcommand\independent{\protect\mathpalette{\protect\independenT}{\perp}}
\def\independenT#1#2{\mathrel{\rlap{$#1#2$}\mkern2mu{#1#2}}}

\newcommand{\R}{\ensuremath{\mathbb{R}}}

\newcommand{\eps}{\varepsilon}  %power set
\newcommand{\1}{\ensuremath{\mathbf{1}}}  %power set
\DeclareMathOperator*{\var}{\text{Var}}
\DeclareMathOperator*{\cov}{\text{Cov}}
 \newcommand\Tau{\mathcal{T}}

\usepackage{amsthm}

\theoremstyle{definition} %plain italicizes stuff

\newtheorem{assumption}{Assumption}

\newtheorem{remark}{Remark}

\DeclareMathOperator*{\argmin}{arg\,min}

\theoremstyle{theorem} 

\newtheorem{lemma}{Lemma}

\newtheorem{assumptiotl}{Assumption}

\newtheorem{assumptioces}{Assumption}

\begin{document}
\bibliographystyle{chicago}

\begin{center}
	
	\quad
	\vspace{5mm}

	\Large{\bf{Flexible estimation of skill formation models}}\symbolfootnote[1]{We thank Gabriella Conti, Janos Gabler, Björn Höppner, Mauricio Olivares and seminar participants at several universities for helpful comments and discussions. The research was supported by the European Research Council (ERC-2020-STG-949319).}

	\vspace{5mm}

	\normalsize 
 	
 	Antonia Antweiler\symbolfootnote[2]{University of Bonn. Email: antonia.antweiler@uni-bonn.de. } \hspace{15mm} Joachim Freyberger\symbolfootnote[3]{University of Bonn. Email: freyberger@uni-bonn.de. } 
 	
	\vspace{5mm}
	
	 \today

	\vspace{1.5cm}
	
\end{center}

\normalsize

\noindent \textbf{Abstract} \medskip

\noindent This paper examines estimation of skill formation models, a critical component in understanding human capital development and its effects on individual outcomes. Existing estimators are either based on moment conditions and only applicable in specific settings or rely on distributional approximations that often do not align with the model. Our method employs an iterative likelihood-based procedure, which flexibly estimates latent variable distributions and recursively incorporates model restrictions across time periods. This approach reduces computational complexity while accommodating nonlinear production functions and measurement systems. Inference can be based on a bootstrap procedure that does not require re-estimating the model for bootstrap samples. Monte Carlo simulations and an empirical application demonstrate that our estimator outperforms existing methods, whose estimators can be substantially biased or noisy.

\vspace{1cm}

\newpage
 
\section{Introduction}

The study of skill formation is central to understanding human capital development and its impact on individual outcomes. The seminal work of \citeN{CH:08} and \citeN{CHS:10} provides a foundational framework for analyzing skill formation as a dynamic, multidimensional process. Building on these papers, a large literature has emerged that explores various aspects of skill formation mechanisms, including persistence, dynamic complementarities, and the optimal timing of investments.\footnote{For example, \citeN{AMN:19} study the interaction of children's cognition and health using data from India, \citeN{ACDMR:19} evaluate the impact of parental investment on socio-emotional and cognitive skills using an RCT in Columbia. \citeN{AJ:16} evaluate the effect of math and verbal skills on educational attainment using data from England. For further references, see \citeN{ACM:2022}, \citeN{AMNS:17}, \citeN{HPS:13}, \citeN{Cunha:11}, \citeN{FK:14}, \citeN{DFW:13} and references therein.} 

Depending on the specification of the production function, estimation in this context can be challenging because the data only contains noisy measures of the latent variables. With the exception of \shortciteN{CHS:10}, early work has therefore primarily employed the Cobb-Douglas production function, which allows parameter estimation using first and second moments of the measures. A more flexible moments-based approach has recently been proposed by \citeN{AW:22}, who use a multi-step instrumental variable (IV) strategy to estimate trans-log production functions. In this iterative procedure, measures replace latent variables in the production function, while other measures serve as instruments. Recent applications of this method include \shortciteN{MFPS:23} and \shortciteN{HRR:24}. 

For general nonlinear production functions, such as the CES production function in \shortciteN{CHS:10}, identification and estimation require the joint distribution of the measures. However, maximum likelihood estimation (MLE) for these models is typically computational prohibitive because the likelihood involves high-dimensional integrals. The dimension of the integral is the product of the number of latent variables and the number of time periods, which results from the need to integrate out latent variables (see Section A5 in their supplement). To reduce the computational burden, \shortciteN{CHS:10} employ nonlinear filtering techniques, which rely on the crucial assumption that the latent variables are (approximately) distributed as mixtures of normals conditional on the measures.  A computationally much simpler and more accessible approach has been suggested by \citeN{AMN:19}, who build on similar approximations but simplify the implementation. To do so, they assume that the joint distribution of all latent variables across time periods is a mixture of normals, which allows estimating the model with a combination of the EM algorithm and nonlinear least squares. Unlike \shortciteN{CHS:10},  whose approach has seen very limited adoption, the method by \shortciteN{AMN:19} has been widely applied, including in studies by  \shortciteN{ACDMR:19}, \shortciteN{ABGN:20}, \shortciteN{ANT:23}, \shortciteN{BFHD:24},  and \citeN{GG:23}.\footnote{A Python implementation of the estimator of \shortciteN{CHS:10} has been provided by Janos Gabler: \href{https://skillmodels.readthedocs.io/}{https://skillmodels.readthedocs.io/}. While very efficient for the class of models considered, it is difficult to adapt to general cases, and it is not directly applicable in the setups of our simulations and our application.}  However, both methods rely on assumptions that may not align with the model. For instance, even if initial skills are normally distributed, nonlinear transformations through CES or trans-log production functions generally result in non-normal distributions in subsequent periods.

Motivated by Monte Carlo simulations, which suggest that the mixture normal approximations of the measures may be poor in some settings, we propose a new estimator of skill formation models. Our approach is based on a likelihood, but uses an iterative procedure to reduce the dimensions of the integrals and thereby the computational complexity. That is, we first flexibly estimate the distribution of the latent variables in the initial period. Subsequently, we proceed recursively: given the distribution of the latent variables in period $t$, we use measures from periods $t$ and $t+1$ to estimate the parameters in period $t$ and the distribution of the latent variables in period $t+1$. While less efficient than full MLE, our method is computationally attractive because each step only estimates parameters of one time period and we integrate out latent variables whose distribution has been estimated in a previous step, which greatly facilitates numerical approximations of the integrals. Inspired by the iterative estimator of \citeN{AW:22}, our procedure generalizes their approach by leveraging all model restrictions across two periods, and it can therefore be used with various flexible specifications of the model, such as different production functions or binary measures. We also provide a bootstrap-based inference method that avoids re-estimating the model for bootstrap samples, enhancing computational efficiency.

We evaluate the properties of our estimator in Monte Carlo simulations and an empirical application. We compare our results with the popular estimators of \shortciteN{AMN:19} (for both the CES and the trans-log production function) and \shortciteN{AW:22} (for the trans-log production function). We find that the performance of the estimator of \shortciteN{AMN:19} depends heavily on the true values of the parameters. In some settings, the bias from the mixture normal approximation is minimal, but in other cases the estimators of the parameters and counterfactuals are severely biased. Increasing the number of mixture components can mitigate the bias, but it may lead to numerical instabilities and larger standard errors. Since our approach is consistent with the specification of the model, it performs well across all scenarios. Compared to \shortciteN{AW:22}, our estimator not only applies to different specifications of the model, but it is also generally more precise, as their method uses only a subset of available moment conditions. Additionally, we demonstrate in the empirical application that their results may depend on which measures are used to replace the latent variables in the production function and which measures are used as instruments.

\textbf{Structure:} Section \ref{s:model} presents the model of skill formation. In Section \ref{s:existing}, we briefly review existing estimators. Section \ref{s:estimation} contains our estimation procedure and the score-based bootstrap. Sections \ref{s:simulations} and \ref{s:application} contain the Monte Carlo simulations and the empirical application, respectively. Finally, Section \ref{s:discussion} provides practical recommendations. 

\section{Model}
\label{s:model}

The description of the model largely follows \citeN{Freyberger:24}. Let $\theta_{t}$ and $I_t$ denote skills and investment at time $t$, respectively. Neither skills nor investment are directly observed and we denote the observed measurements by $Z_{\theta,t,m}$ and $Z_{I,t,m}$, respectively. We consider a model based on
\begin{eqnarray}
\ln\theta_{t+1} &=& f(\ln\theta_{t},\ln I_{t},\delta_{t}) + \eta_{\theta,t} 
\hspace{34mm} t = 0, \ldots, T-1 \label{eq:prod_fn} \\
Z_{\theta,t,m} &=& \mu_{\theta,t,m} + \lambda_{\theta,t,m} \ln \theta_{t} + \eps_{\theta,t,m} \hspace{26mm} t = 0, \ldots, T, m = 1,2,3 \label{eq:measurement_eq_skills} \\
Z_{I,t,m} &=& \mu_{I,t,m} + \lambda_{I,t,m} \ln I_{t} + \eps_{I,t,m} \hspace{26mm} t = 0, \ldots, T-1, m = 1,2,3. \label{eq:measurement_eq_invest}  
\end{eqnarray}
The first equation describes the production technology with a production function $f$ that depends on skills and investment at time $t$, a parameter vector $\delta_{t}$, and an unobserved shock $\eta_{\theta,t}$. We assume an additive error mainly for expositional purposes and because that restriction is commonly used in parametric specifications. The second and third equation describe the measurement system for unobserved (latent) skills $\theta_{t}$ and unobserved investment $I_t$, respectively. Observed investment is a special case with $ \mu_{I,t,m} = 0$, $\lambda_{I,t,m} = 1$, and $\eps_{I,t,m} = 0$ for all $m$ and $t$, in which case $Z_{I,t,m} = \ln I_{t}$. We discuss alternative specifications of the measurement system in Section \ref{s:discussion}.

Next, we introduce two equations to allow for endogenous investment and anchoring at an adult outcome. If investment is exogenous, in the sense that $\eta_{\theta,t}$ is independent of $I_{t}$, then these equations are not required for the main results. However, modeling investment explicitly may still be useful as it allows studying certain counterfactuals, such as the effect of changes in income on skills. In particular, we let
\begin{eqnarray}
	\ln I_t &=& \beta_{0t} + \beta_{1t} \ln \theta_{t}  + \beta_{2t} \ln Y_{t}   + \eta_{I,t} \hspace{26mm} t = 0, \ldots, T-1  \label{eq:investment}   \\
	Q &=& \rho_{0} + \rho_{1} \ln \theta_{T} + \eta_Q.\label{eq:anchor_eq}
\end{eqnarray}
Here,  $Y_t$ is parental income (or another exogenous variable that affects investment) and $Q$ is an adult outcome, such as earnings or education. An adult outcome does not necessarily have to be available and we can simply use a skill measure in period $T$ in its place.

In summary, the observed variables are income $\{Y_t\}_{t=0}^{T-1}$, the measures  $\{Z_{\theta,t,m}\}_{t=0,\ldots,T, m= 1,2,3}$ and $\{ Z_{I,t,m} \}_{t=0,\ldots,T-1, m= 1,2,3}$,  and the adult outcome $Q$, but we neither observe skills $\{\theta_{t}\}^{T}_{t=0}$ nor investment  $\{I_t\}_{t=0}^{T-1}$. We also do not observe any of the errors/shocks in the five equations. The parameters of interest are $\{\mu_{\theta,t,m},\lambda_{\theta,t,m}\}_{t=0,\ldots,T,m=1,2,3}$, $\{\mu_{I,t,m},\lambda_{I,t,m}\}_{t=0,\ldots,T-1,m=1,2,3}$, $\{\delta_t\}^{T-1}_{t=0}$, $\{ \beta_{0t}, \beta_{1t}, \beta_{2t}\}^{T-1}_{t=0}$, $(\rho_0,\rho_1)$, and the distributions of the latent variables.

In the following analysis, we mainly consider the two most commonly used forms for the production technology in the empirical literature, namely the trans-log production function with
	\begin{equation}
		\label{eq:translog}
	\ln \theta_{t+1} = a_t + \gamma_{1t}\ln \theta_{t} + \gamma_{2t} \ln I_{t} + \gamma_{3t}\ln \theta_{t} \ln I_{t}  + \eta_{\theta,t} 
  \end{equation}
	 and parameter vector $\delta_{t} = (a_t, \gamma_{1t},\gamma_{2t},\gamma_{3t})$ and the CES production function with
	\begin{equation}
	\label{eq:ces}
	\theta_{t+1} =  ( \gamma_{1t} \theta_{t}^{\sigma_{t}} + \gamma_{2t} I_{t}^{\sigma_{t}} )^{\psi_t/\sigma_{t}} \exp(\eta_{\theta,t})
  \end{equation}
	 and parameter vector $\delta_{t} = (\gamma_{1t},\gamma_{2t},\sigma_{t},\psi_t)$, where $\gamma_{1t},\gamma_{2t},\psi_t,\sigma_{t} \neq 0$. When $\sigma_{t} = 0$, the CES reduces to the Cobb-Douglas production function. 
	 
We now state several additional assumptions that are common in the literature.

\begin{assumption}\label{a:baseline}  \qquad 
	
	\begin{enumerate}[(a)]
		
		\item $\{\{\eps_{\theta,t,m}\}_{t=0,\ldots,T, m= 1,2,3}, \{\eps_{I,t,m} \}_{t=0,\ldots,T-1, m= 1,2,3}, \eta_Q\}$ are jointly independent and independent of $\{\{\theta_{t}\}^{T}_{t=0},\{I_t\}_{t=0}^{T-1}\}$ conditional on $\{Y_t\}_{t=0}^{T-1}$.

		\item All random variables have bounded first and second moments.

		\item $E[\eps_{\theta,t,m}] = E[\eps_{I,t,m}] = E[\eps_Q] =  0$ for all $t$ and $m$.
		  
		\item $\lambda_{\theta,t,m}, \lambda_{I,t,m} \neq 0$ for all $t$ and $m$.

		\item For all $t$ and $m$, the real zeros of the characteristic functions of $\eps_{\theta,t,m}$ are isolated and distinct from those of its derivatives. Identical conditions hold for the characteristic functions of $\eps_{I,t,m}$ and $\eta_{Q}$.
		
		\item The support of $(\theta_{t},I_{t},Y_t)$ includes an open ball in $\R^3$ for all $t$.

		\item  $\eta_{I,t} \independent (\theta_{t},Y_t)$ and $E[\eta_{I,t}]= 0$ for all $t$. Moreover,  $\eta_{\theta,t} = \kappa_t \eta_{I,t} + \eps_{C,t}$ where $\kappa_t$ is a constant, $\eps_{C,t} \independent (\theta_{t},Y_t,I_t)$, and $E[\eps_{C,t} ] = 0$ for all $t$.

	\end{enumerate}
	 
\end{assumption}

Part (a) imposes common independence assumptions on the measurement errors. Importantly, $I_t$ and $\theta_t$ are not independent and $I_t$ may be endogenous and contemporaneously correlated with $\eta_{\theta,t}$. Part (b) is a standard restriction, part (c) is needed because all measurement equations contain an intercept, and part (d) ensures that the skills actually affect the measures. Part (e) contains weak regularity conditions needed for nonparametric identification of the distributions of skills and investment and that hold for most common distributions.    Part (f) is a mild support condition that ensures sufficient variation of $(\theta_{t},I_{t},Y_t)$. Part (g) allows for endogenous investment (i.e. $\eta_{\theta,t} \not\independent I_t$), but implies that $Y_t$ can serve as an instrument with identification based on a control function argument, as in \shortciteN{AMN:19}. Exogenous investment is a special case with $\kappa_{t} = 0$.  

Notice that we assume that exactly three measures are available for skills and investment in each period. It is straightforward to allow for more measures at the expense of additional notation. Identification can also be achieved with two measures. In this case, one either needs additional assumption on the correlation between skills and investment across time periods (see Assumption 1(e) in \citeN{Freyberger:24}) or use specific functional forms for the production function (as discussed below).  Some of the other assumptions, such as parts (a) and (g), are also stronger than necessary for point identification depending on which production function is employed. For instance, with the Cobb-Douglas production function, identification can be achieved based on the first two moments, as discussed below. We use these stronger assumptions because they greatly simplify the likelihood.

In addition to Assumption \ref{a:baseline}, we impose scale and location restrictions, which are necessary for point identification of the parameters of the model. As explained in \citeN{Freyberger:24}, there are different ways how these restrictions can be imposed, the specific choice affects the estimated parameters, but many relevant summary statistics and counterfactuals are invariant to these restrictions and identified without them. We focus on those features in our simulations and in the empirical application. Therefore, we only state one set of assumptions and refer the reader to \citeN{Freyberger:24} for alternative choices.

\setcounter{assumptiotl}{1}
 \setcounter{assumptioces}{1}

With the trans-log production function in Equation (\ref{eq:translog}), we use the following restrictions
\begin{assumptiotl}\label{a:normalizations_tl}  \qquad 
\begin{itemize}
    \item $\mu_{\theta,t,1} = 0$ and $\lambda_{\theta,t,1} = 1$ for all $t = 0, \ldots, T$  
\item $\mu_{I,t,1} = 0$ and $\lambda_{I,t,1} = 1$ for all $t = 0, \ldots, T-1$.
\end{itemize}
\end{assumptiotl}
\noindent With the CES production in Equation (\ref{eq:ces}), we impose 
\begin{assumptioces}\label{a:normalizations_ces}  \qquad 
\begin{itemize} 
\item $\mu_{\theta,t,1} = 0$ for all $t = 0, \ldots, T$  
\item $\mu_{I,t,1} =  0$ for all $t = 0, \ldots, T-1$  
\item $\psi_{t} = 1$ for all $t=0,\ldots,T-1$
\item $\lambda_{\theta,0,1}=1.$
\end{itemize}
\end{assumptioces}
If we allowed for $\psi_{t} \neq 1$, we would have to impose additional restrictions on the measurement system, namely $\lambda_{\theta,t,1} = \lambda_{\theta,t+1,1}$ for all $t = 0, \ldots, T-1$. Either case would yield the same summary statistics and counterfactuals we report below. We focus on $\psi_{t} = 1$ because it is a common restriction in empirical applications. Notice that in this case, we only impose one scale restriction in one time period (i.e. $\lambda_{\theta,0,1}=1$) as opposed to multiple scale restrictions in multiple time periods in the trans-log case. 

The following lemma states that the model is identified under the previous assumptions. The proof follows from Corollaries 1 and 2 of \citeN{Freyberger:24}.

\begin{lemma}
    Suppose Assumption \ref{a:baseline} holds. 
    
    \begin{enumerate}
        \item[(a)] With a trans-log production function as in Equation (\ref{eq:translog}) and under Assumption \ref{a:normalizations_tl}, all parameters and error distributions in Equations (\ref{eq:prod_fn})--(\ref{eq:anchor_eq}) are point identified. 
\item[(b)] With a CES production function as in Equation (\ref{eq:ces}) and under Assumption \ref{a:normalizations_ces}, all parameters and error distributions in Equations (\ref{eq:prod_fn})--(\ref{eq:anchor_eq}) are point identified.
        
    \end{enumerate} 
\end{lemma}
 
\section{Existing estimators}
\label{s:existing}

In this section, we briefly describe existing estimators that have been used for this class of models. We first consider $\kappa_t = 0$ in Assumption \ref{a:baseline} and then discuss how endogenous investment can be allowed for afterwards.

\subsection{Estimators for the Cobb-Douglas production function}
\label{s:cobb-douglas}

In the special case where the trans-log production function simplifies to the Cobb-Douglas production function, the parameters of the model can be identified using the first two moments of the measurements, as in \citeN{CH:08}. To see why, write the production function as 
$$\ln \theta_{t+1} = a_t + \gamma_{1t}\ln \theta_{t} + \gamma_{2t} \ln I_{t}  + \eta_{\theta,t} $$
and notice that 
$$ \begin{pmatrix}
\gamma_{1t} \\ \gamma_{2t} 
\end{pmatrix}
= 
\begin{pmatrix}
\var(\ln \theta_{t}) & \cov(\ln \theta_{t},\ln I_{t} ) \\  \cov(\ln \theta_{t},\ln I_{t} ) &  \var(\ln I_{t})
\end{pmatrix}^{-1}
\begin{pmatrix}
\cov(\ln \theta_{t},\ln \theta_{t+1}) \\ \cov(\ln I_{t},\ln \theta_{t+1}) 
\end{pmatrix}.
$$
Moreover, it follows from Assumptions \ref{a:baseline} and \ref{a:normalizations_tl} that $\cov(\ln \theta_{t},\ln I_{t}) = \cov(Z_{I,t,1},Z_{\theta,t,1}) $ and $\cov((\ln \theta_{t},\ln I_{t})',\ln \theta_{t+1}) = \cov((Z_{\theta,t,1},Z_{I,t,1})',Z_{\theta,t+1,1})$. Finally, it is well known that $\var(\ln \theta_{t})$ and $\var(\ln I_{t})$ are identified with three measures in each period using standard arguments from linear factor models (going back to \citeN{AR:56} or \citeN{Madansky:64}). Using similar arguments, all other parameters are identified as well. These identification arguments are constructive and allow for sample analog estimation. 

An alternative approach, used by \citeN{HPS:13} and \citeN{ACDMR:19}, is to first construct individual estimates of skills and investment using the so-called Bartlett scores. Using the population parameters, these scores for log-skills are 
$$\widehat{\ln \theta}_t = \sum^M_{m=1} w_{\theta,t,m} Z_{\theta,t,m} \quad \text{with} \quad w_{\theta,t,m} = \left(\sum^M_{m=1} \lambda_{\theta,t,m}^2/\var(\eps_{\theta,t,m}) \right)^{-1}  \left(\lambda_{\theta,t,m}/\var(\eps_{\theta,t,m}) \right).$$ 
For example, if $\lambda_{\theta,t,m} = 1$ for all $m$ and $\var(\eps_{\theta,t,m})$ is identical for all $m$, then the Bartlett score simplifies to $\widehat{\ln \theta}_t = \frac13\sum^M_{m=1} Z_{\theta,t,m} = \ln \theta_t +  \frac13\sum^M_{m=1} \eps_{\theta,t,m}$. While the Bartlett score is an unbiased estimator of $\ln \theta_t$, it is still subject to measurement error. Moreover, in practice, the weights have to be estimated. Hence, using the estimated scores instead of the true latent variables in a regression to estimate the production function parameters yields biased and inconsistent estimators of the parameters. However, the (large sample) bias only depends on the first two moments of the data and can therefore be estimated. The resulting bias corrected estimator uses the same identifying assumptions and is based on very similar moment conditions as the approach of \citeN{CH:08}.

For certain counterfactuals, one has to additionally identify and estimate the joint distribution of skills and investment across all time periods. One way to do so is to first identify and estimate the joint distribution in the initial time period, which in practice is typically achieved using distributional assumptions. One could then impose parametric assumptions on $\eta_{\theta,t}$ and $\eta_{I,t}$ (e.g. assuming that they are normally distributed) and use the recursive structure of the model to identify the joint distribution of all latent variables.

\subsection{Estimator of Agostinelli and Wiswall (2025)}
\label{s:aw}

Another interpretation of the moment-based estimator is an instrumental variable approach. With the Cobb-Douglas production function, we can replace $\ln \theta_{t+1}$, $\ln \theta_{t}$, and $\ln I_{t}$ in the production function with their first measures (for which the scale and location restrictions are imposed) to obtain
$$Z_{\theta,t+1,1} = a_t + \gamma_{1t}Z_{\theta,t,1} + \gamma_{2t}  Z_{I,t,1}   + \eta_{\theta,t} - \gamma_{1t}\eps_{\theta,t,1}   - \gamma_{2t}\eps_{I,t,1}  + \eps_{\theta,t+1,1}. $$
Next, notice that due to the measurement errors,  $Z_{\theta,t,1}$ and $Z_{I,t,1} $ are endogenous, but  $Z_{\theta,t,2}$ and $Z_{I,t,2} $ are valid and relevant instruments under Assumption \ref{a:baseline}. Notice that this approach highlights that only two measures for each latent variable are required.  Intuitively, the reason is that in a regression of $\ln \theta_{t+1}$ on $\ln \theta_{t}$, the slope coefficient is $$\frac{\cov(\ln \theta_{t+1},\ln \theta_{t})}{\var(\ln \theta_{t})} = \frac{\lambda_{\theta,t,2} \cov(\ln \theta_{t+1},\ln \theta_{t})}{\lambda_{\theta,t,2} \var(\ln \theta_{t})} 
 = \frac{\cov(Z_{\theta,t+1,1} ,Z_{\theta,t,2} )}{\cov(Z_{\theta,t,1} ,Z_{\theta,t,2} )}.$$
Hence, the slope coefficient is identified with two measures, even though $\cov(\ln \theta_{t+1},\ln \theta_{t})$ and $\var(\ln \theta_{t})$ are not separately identified in this case.

Having three measures has the advantage that the parameters of the measurement system are identified as well. One can then define
$$\tilde{Z}_{\theta,t,m} = \frac{Z_{\theta,t,m} - \mu_{\theta,t,m}}{\lambda_{\theta,t,m}} \qquad \text{ and } \qquad \tilde{Z}_{I,t,m} = \frac{Z_{I,t,m} - \mu_{I,t,m}}{\lambda_{I,t,m}}  $$
and use any of these variables to replace  $\ln \theta_{t+1}$, $\ln \theta_{t}$, and $\ln I_{t}$ in the production function. All other measures then serve as valid instruments. 

\citeN{AW:22} extent this approach to the trans-log production function. In that case, the product $\ln \theta_{t}\ln I_{t}$ is replaced by $\tilde{Z}_{\theta,t,m}\tilde{Z}_{I,t,m}$ and products of the measures are used as additional instruments. Such an estimation procedure requires that the production function is linear in the parameters and it is therefore not directly extendable to nonlinear production functions, such as the CES. Moreover, although not discussed by \citeN{AW:22}, there is a potentially large set of moment conditions, because past and future measures are also valid instruments and any of the measures can be used to replace a latent variable. Combining all moments efficiently is an interesting open question.   

As discussed in Section \ref{s:cobb-douglas} (and as done in \citeN{AW:22}), one can impose additional assumptions to identify the joint distribution of skills and investment after identifying the production function parameters. A nice feature of this approach is that is does not require distributional assumptions on the measurement errors $\eps_{\theta,t,m}$ and $\eps_{I,t,m}$ for $t>0$.
 
\subsection{Estimator of Cunha, Heckman, and Schennach (2010)}
\label{s:chs}

\shortciteN{CHS:10} estimate the model based on a maximum likelihood approach, which has the advantage of using all available information. The likelihood function in this context is very complex because one has to integrate out the latent variables, and high-dimensional integrals are costly to compute numerically.  

To circumvent this problem, \shortciteN{CHS:10} approximate the density of the latent variables at time $t+k$ for $k\in \{0,1\}$ conditional on all measures up to time $t$ with a normal mixture distribution. To better explain this approximation, consider $t=0$ and $k=1$. Moreover, assume that in the initial period the distribution of $\ln \theta_0$ is a mixture of $L$ normals. That is,
$$f_{\ln \theta_0}(s) = \sum^L_{l=1} p_l f_l(s)$$
where $f_l$ for $l= 1, \ldots, L$ are the component distributions and $p_l$ are the weights. Also, assume that $\eps_{\theta,0,m}$ is standard normally distributed for all $m$. Due to independence of the measurement error, we then have
$$f_{Z_{\theta,0,1},Z_{\theta,0,2},Z_{\theta,0,3} \mid \ln \theta_0} (z_1,z_2,z_3 \mid s)  = \prod^3_{m=1} \phi( z_1 - \mu_{\theta,0,m} - \lambda_{\theta,0,m} s),$$
where $\phi$ denotes the standard normal pdf. Consequently,
$$f_{Z_{\theta,0,1},Z_{\theta,0,2},Z_{\theta,0,3}, \ln \theta_0} (z_1,z_2,z_3,s)  = \sum^L_{l=1} p_l\prod^3_{m=1}  f_l(s)  \phi( z_m - \mu_{\theta,0,m} - \lambda_{\theta,0,m} s).$$
Notice that by  the properties of the normal distribution, we can write
$$\prod^3_{m=1}  f_l(s)  \phi( z_m - \mu_{\theta,0,m} - \lambda_{\theta,0,m} s) = g_l(z_1,z_2,z_3,s) $$
where $g_l$ is a joint pdf of a four-dimensional normal random vector. Also define
$$g_l(z_1,z_2,z_3) = \int \prod^3_{m=1}  f_l(s)  \phi( z_m - \mu_{\theta,0,m} - \lambda_{\theta,0,m} s) ds$$
which are the corresponding marginal distributions of the measures. It follows that
\begin{align*}
f_{\ln \theta_0 \mid Z_{\theta,0,1},Z_{\theta,0,2},Z_{\theta,0,m}} ( s \mid z_1,z_2,z_3)  &= \frac{\sum^L_{l=1} p_l g_l(z_1,z_2,z_3,s)}{f_{Z_{\theta,0,1},Z_{\theta,0,2},Z_{\theta,0,m}}(z_1,z_2,z_3)}    \\
&= \frac{\sum^L_{l=1} p_l g_l(z_1,z_2,z_3,s)}{\sum^L_{l=1} p_l g_l(z_1,z_2,z_3)}    \\
&=  \sum^L_{l=1} \tilde{p}_l(z_1,z_2,z_3)  \frac{g_l(z_1,z_2,z_3,s)}{g_l(z_1,z_2,z_3)}
\end{align*}
where
$$\tilde{p}_l(z_1,z_2,z_3) = \frac{p_l g_l(z_1,z_2,z_3)}{\sum^L_{l=1}  p_l g_l(z_1,z_2,z_3) }. $$
Hence, $\ln \theta_0$ has a mixture normal distribution conditional on $(Z_{\theta,0,1},Z_{\theta,0,2},Z_{\theta,0,3})$. Both the weights and the component distributions depend on the measures. It can also be shown using similar assumptions and arguments that $\ln I_{0}$ has a mixture normal distribution conditional on its measures under similar assumptions.

As mentioned above, \shortciteN{CHS:10} approximate the distribution of $\ln \theta_1$, conditional on the measures in period $0$, with a mixture of normals as well. To analyze this approximation, first consider the Cobb-Douglas production function 
where
$$\ln \theta_{1} = a_0 + \gamma_{10}\ln \theta_{0} + \gamma_{20} \ln I_{0} +  \eta_{\theta,0}. $$
In this case, if the latent variables in period $0$ both have a mixture normal distribution conditional on the measures in period 0 and if $\eta_{\theta,0}$ is normally distributed, then the approximation of \shortciteN{CHS:10} is in fact exact, because a linear combination of normally distributed random variables is also normally distributed.

However, with the CES production, we have
$$\theta_{1} =  ( \gamma_{10} \theta_{0}^{\sigma_{0}} + \gamma_{20} I_{0}^{\sigma_{0}} )^{\psi_0/\sigma_{0}} \exp(\eta_{\theta,0})$$
and a nonlinear function of normal random variables are generally not normally distributed. In particular, we can write
$$\theta_{1}^{\sigma_{0}/\psi_0}  =  ( \gamma_{10} \bar{\theta}_0 + \gamma_{20} \bar{I}_0 )$$
where 
$$ \bar{\theta}_0 =   \theta_{0}^{\sigma_{0}}\exp((\sigma_{0}/\psi_0)\eta_{\theta,0}) \quad \text{and}   \quad \bar{I}_0 = I_{0}^{\sigma_{0}}\exp((\sigma_{0}/\psi_0)\eta_{\theta,0}).$$ 
It is easy to show that if $\eta_{\theta,0}$ is normally distributed, then $\ln \bar{\theta}_0$  and $\ln  \bar{I}_0$  have a joint mixture normal distribution conditional on the  measures. The conditional distribution of $ \bar{\theta}_0$  and $  \bar{I}_0$ is therefore a mixture of log-normals. For the approximation of \shortciteN{CHS:10} to be exact in this case, it would have to hold that a linear combination of log-normal random variables is also log-normal, which is not the case. Hence, with the CES production function, the estimator of \shortciteN{CHS:10} is generally inconsistent. The approximation can be accurate if the number of mixture components is large, which comes at the expense of computational complexity and potentially large standard errors. Even with a small number of mixtures (\shortciteN{CHS:10} use two components in their application), such an approximation may be quite precise \cite{HB:15}, but the quality of the approximation depends heavily on the true parameters.

\subsection{Estimator of Attanasio, Meghir, and Nix (2020) }
\label{s:amn}

The only other published paper we are aware of that has employed the approach of \shortciteN{CHS:10} is \citeN{Pavan:06}. More recently, \shortciteN{AMN:19} suggested an alternative estimator that relies on similar approximations as \shortciteN{CHS:10}, but is much easier to implement. This estimator has since been used in several papers, including \shortciteN{AMNS:17}, \shortciteN{ABGN:20},   \shortciteN{ANT:23}, \citeN{GG:23}, and \shortciteN{BFHD:24}.

\shortciteN{AMN:19} approximate the joint distribution of all latent log-skills and log-investment using a mixture of normals. Assuming that the measurement errors $\eps_{\theta,t,m}$ and $\eps_{I,t,m}$ as well as income are also normally distributed, it follows that the joint distribution of all observed variables in all time periods is a mixture of normals. \shortciteN{AMN:19} estimate the model in three steps. First, they estimate the joint distribution of all observed variables. Second, they use the factor structure to estimate the distribution of all latent variables by minimum distance. Third, they take draws from that distribution and estimate the production and investment function parameters by nonlinear least squares. The papers above, that have employed this approach, all use a mixture of at most two normals.

While the approximation of \shortciteN{AMN:19} is different compared to \shortciteN{CHS:10}, they have a similar flavor. They are both exact in certain parametric specifications with a Cobb-Douglas production function. However, since a linear combination of log-normals is not log-normally distributed, both approximations are not exact and may be poor with the CES production function.

As pointed out by \citeN{Freyberger:24}, \shortciteN{AMN:19} ``normalize''  parameters of the measurement system that are in fact identified with the CES production function. Hence, if these parameters are not set to the (unknown) true values, the estimator is inconsistent. \citeN{Freyberger:24} also provides an adapted estimator, which estimates all identified parameters. We rely on that estimator in our simulations and in the application.

\subsection{Endogenous investment}
\label{s:endo_invest}

If $\kappa_t \neq 0$ in Assumption \ref{a:baseline}, one can use a control function approach. To do so, notice that 
\begin{align*}
\eta_{I,t} & =  \ln I_t - E[\ln I_t \mid \ln \theta_{t} ,\ln Y_{t} ] \\
& =  \ln I_t - \beta_{0t} - \beta_{1t} \ln \theta_{t}  - \beta_{2t} \ln Y_{t}     
\end{align*}
which can be estimated using, for example, a moment-based approach as in Section \ref{s:cobb-douglas}. Using part (g) of Assumption \ref{a:baseline}, we can write
\begin{align*}
\ln\theta_{t+1}  &= f(\ln\theta_{t},\ln I_{t},\delta_{t}) + \eta_{\theta,t} \\
 &= f(\ln\theta_{t},\ln I_{t},\delta_{t}) +   \kappa_t \eta_{I,t} + \eps_{C,t}.
\end{align*}
Hence, once we include $\eta_{I,t}$ as an additional covariate in the production function, all inputs are exogenous, and we can estimate the parameters using one of the previously discussed approaches.

\section{Estimation and inference}
\label{s:estimation}

Maximum likelihood has two main advantages in this set-up. First, since the model is nonparametrically identified \shortcite{CHS:10,Freyberger:24}, a maximum likelihood estimator allows for flexible parametric specifications of the production function and the measurement system, whereas moment-based estimation is feasible only in specific settings. Second, a MLE efficiently uses all assumptions imposed, such as all independence conditions. Maximum likelihood also has two disadvantages compared to moment-based estimators. First, it requires a full parametric model, even if the production function parameters might be estimable using moment conditions only. However, notice that counterfactuals often require the joint distribution of skills and investment, in which case additional parametric assumptions are also needed with moment-based estimation (as in \citeN{AW:22}). Second, MLE using all time periods without approximations as in \shortciteN{CHS:10} and \shortciteN{AMN:19} is computationally prohibitive.

We suggest a new maximum likelihood approach that solves the computational challenge by using an iterative procedure. That is, we first estimate the distribution of the latent variables in the initial period. We then proceed recursively: Given the distribution of the latent variables in period $t$, we use measures from periods $t$ and $t+1$ to estimate the parameters in period $t$ and the distribution of the latent variables in period $t+1$. Even though the distributions of the latent variables might not be available in closed form, we can simulate from them to calculate the likelihood. Our iterative procedure is computationally feasible and it is similar to \citeN{AW:22}, who use an iterative moment-based procedure with the trans-log production function. Compared to the (computationally prohibitive) full MLE, our estimator is less efficient as it does not combine data from all time periods.

In the next two subsections, we explain our estimator in more detail and describe a computationally attractive bootstrap procedure, which does not require re-estimating the model for the bootstrap samples. We explain our estimator in general terms, using Equations (\ref{eq:prod_fn})--(\ref{eq:anchor_eq}) with exogenous investment for notational convenience. 

In Monte Carlo simulations and the empirical application, we compare our estimator to those of \shortciteN{AMN:19} and \citeN{AW:22}, which are widely used in applications with the CES and trans-log production function, respectively.

\subsection{Likelihood and estimator}
\label{s:likelihood}

To simplify the notation, define $Z_{\theta,t} = (Z_{\theta,t,1},Z_{\theta,t,2},Z_{\theta,t,3})'\in \R^3$ for all $t$ and define analogously $Z_{I,t}$ for all $t$. In the first step, we estimate the conditional distribution of $\ln \theta_0 \mid \ln Y_0$. To do so, note that by the law of total probability, we can write
\begin{align*}
       f_{Z_{\theta,0} \mid \ln Y_0 }  &= \int  f_{ Z_{\theta,0},\ln \theta_0 \mid  \ln Y_0  }  d \ln \theta_0 \\
     &= \int  f_{ Z_{\theta,0} \mid \ln Y_0, \ln \theta_0 }  f_{ \ln \theta_0 \mid \ln Y_0} d \ln \theta_0 \\
     &= \int \prod^3_{m=1} f_{ Z_{\theta,0,m} \mid \ln Y_0, \ln \theta_0 }  f_{ \ln \theta_0 \mid \ln Y_0} d \ln \theta_0. 
\end{align*} 
More precisely, using the linear measurement system and $z_{\theta,t} = (z_{\theta,t,1},z_{\theta,t, 2},z_{\theta,t,3})'$, we get
\begin{align*}
      f_{Z_{\theta,0} \mid \ln Y_0 }(z_{\theta, 0} \mid y  ) 
     = \int \prod^3_{m=1} f_{ \eps_{\theta,0,m}}(z_{\theta,0,m} - \mu_{\theta,0,m} - \lambda_{\theta,0,m} q ) f_{ \ln \theta_0 \mid \ln Y_0}(q | y)  d q 
\end{align*} 
where we make use of independence of the measurement errors. We can now use a flexible parametric specification for $ f_{ \ln \theta_0 \mid \ln Y_0}$, such as a mixture of normals, and estimate that distribution along with $f_{ \eps_{\theta,0,m}}$ and the parameters of the measurement system by MLE.

Next, we use the skill measures from periods $0$ and $1$ and the investment measures from period $0$ and write
\begin{align*}
     f_{Z_{\theta,0}, Z_{I,0} ,Z_{\theta,1} \mid \ln Y_0 }&  = \iint   f_{Z_{\theta,0}, Z_{I,0} ,Z_{\theta,1}\mid \ln Y_0, \ln \theta_0, \ln I_0 }     f_{\ln I_0,  \ln \theta_0 \mid \ln Y_0} d \ln \theta_0  d \ln I_0  \\
     &= \iint \prod^3_{m=1}f_{Z_{\theta,0,m} \mid  \ln \theta_0 } f_{Z_{I,0,m} \mid  \ln I_0 } f_{Z_{\theta,1} \mid \ln Y_0, \ln \theta_0, \ln I_0 }   f_{\ln I_0,\ln \theta_0 \mid \ln Y_0} d \ln \theta_0  d \ln I_0 
     \end{align*} 
where we again make use of independence of the measurement errors. Now, notice that
\begin{align*}
    Z_{\theta,1,m} &=  \mu_{\theta,1,m} + \lambda_{\theta,1,m} \ln \theta_{1} + \eps_{\theta,1,m} \\
    &= \mu_{\theta,1,m} + \lambda_{\theta,1,m} f(\ln \theta_{0}, \ln I_{0},\delta_{0}) + \lambda_{\theta,1,m} \eta_{\theta,0}  + \eps_{\theta,1,m}.
\end{align*} 
If we, additionally to $\ln \theta_0$ and $\ln I_0$, also condition on the production function shock $\eta_{\theta,0}$, we can again make use of independence of the measurement errors to conclude that the different measures are conditionally independent. Hence, we can write the joint distribution of the skill measures in $t=1$ as the product of their marginals, so that
\begin{align*}
     & f_{Z_{\theta,0}, Z_{I,0} ,Z_{\theta,1} \mid \ln Y_0 }\\
     & \quad = \iiint \prod^3_{m=1}f_{Z_{\theta,0,m} \mid  \ln \theta_0 } f_{Z_{I,0,m} \mid  \ln I_0 } f_{Z_{\theta,1,m}  \mid \ln Y_0,  \ln \theta_0, \ln I_0, \eta_{\theta,0} }  f_{\ln I_0,  \ln \theta_0, \eta_{\theta,0} \mid \ln Y_0} d \ln \theta_0  d \ln I_0 d \eta_{\theta,0}.
 \end{align*} 
We can further write the joint density of $\ln I_0,  \ln \theta_0, \eta_{\theta,0} \mid \ln Y_0$ as
\begin{align*}
    f_{\ln I_0,  \ln \theta_0, \eta_{\theta,0} \mid \ln Y_0} = f_{\eta_{\theta,0} \mid \ln I_0,  \ln \theta_0,  \ln Y_0} f_{\ln I_0 \mid \ln \theta_0,  \ln Y_0} f_{ \ln \theta_0 \mid  \ln Y_0}. 
\end{align*}
All components of the likelihood now have simple expressions in terms of the model. Let $z_{I,t,m} \in \R$ for all $t$ and $m$. Assuming exogenous investment, we get
\begin{align*}
    f_{Z_{\theta,0,m} \mid  \ln \theta_0 }(z_{\theta,0,m} \mid q) &=   f_{ \eps_{\theta,0,m}}(z_{\theta,0,m} - \mu_{\theta,0,m} - \lambda_{\theta,0,m} q ) \\
    f_{Z_{I,0,m} \mid  \ln I_0 }(z_{I,0,m} \mid i) &=   f_{ \eps_{I,0,m}}(z_{I,0,m} - \mu_{I,0,m} - \lambda_{I,0,m} i ) \\
       f_{Z_{\theta,1,m}  \mid \ln Y_0,  \ln \theta_0, \ln I_0, \eta_{\theta,0} }(z_{\theta,1,m} \mid y, q, i, e) &=   f_{ \eps_{\theta,1,m}}(z_{\theta,1,m} -\mu_{\theta,1,m} - \lambda_{\theta,1,m} f( q, i,\delta_{0}) - \lambda_{\theta,1,m} e ) \\
       f_{\eta_{\theta,0} \mid \ln I_0,  \ln \theta_0,  \ln Y_0}(e \mid i,q,y) &= f_{\eta_{\theta,0}}(e) \\
       f_{\ln I_0 \mid \ln \theta_0,  \ln Y_0}(i \mid q, y) &=f_{ \eta_{I,0}}(i - (\beta_{00} + \beta_{10} q  + \beta_{20}  y)).
\end{align*}
Also, notice that $f_{Z_{\theta,0,m} \mid  \ln \theta_0 }$ and $f_{ \ln \theta_0 \mid  \ln Y_0}$ have already been estimated in the initial step. Hence, the only remaining parameters are the parameters in the measurement error equations for skills in period $1$ and for investment in period $0$, the production function, the investment equation in period $0$ and the corresponding error distribution in period $0$.

We can now simulate draws from the initial distribution $\ln \theta_0 \mid \ln Y_0$ and from the investment error distribution $\eta_{I,0}$. Equipped with those draws and the estimated investment equation parameters, we generate $\ln I_0$. Now, we proceed by taking draws from the production function shock distribution to generate $\ln \theta_1$ with the estimates from the production function. We have a constructed a synthetic sample $\ln \theta_1 \mid \ln Y_0$. Based on those draws, we can nonparametrically estimate the density function (arbitrarily well given the estimated parameters).  

We now analyze the joint distribution of skill measures in periods $1$ and $2$ and investment measures in period 1, conditional on income in periods $0$ and $1$. Following previous arguments, we have 
\begin{align*}
      &  f_{Z_{\theta,1}, Z_{I,1}, Z_{\theta,2} \mid \ln Y_0, \ln Y_1 } \\
      & \; = \iiint \prod^3_{m=1}f_{Z_{\theta,1,m} \mid  \ln \theta_1 } f_{Z_{I,1,m} \mid  \ln I_1 } f_{Z_{\theta,2,m}  \mid \ln Y_0, \ln Y_1,  \ln \theta_1, \ln I_1, \eta_{\theta,1} }  f_{\ln I_1,  \ln \theta_1, \eta_{\theta,1} \mid \ln Y_0, \ln Y_1} d \ln \theta_1  d \ln I_1 d \eta_{\theta,1}\\
      & \; = \iiint \prod^3_{m=1}f_{Z_{\theta,1,m} \mid  \ln \theta_1 } f_{Z_{I,1,m} \mid  \ln I_1 } f_{Z_{\theta,2,m}  \mid \ln \theta_1, \ln I_1, \eta_{\theta,1} } f_{\eta_{\theta,1}} f_{\ln I_1 \mid \ln \theta_1,  \ln Y_1} f_{ \ln \theta_1 \mid  \ln Y_0} d \ln \theta_1  d \ln I_1 d \eta_{\theta,1}.
\end{align*} 
where we use the simplifications  $f_{\ln \theta_1 \mid \ln Y_0, \ln Y_1} = f_{\ln \theta_1 \mid \ln Y_0}$, $ f_{\ln I_1 \mid \ln \theta_1, \ln Y_0, \ln Y_1 }=  f_{\ln I_1 \mid \ln \theta_1, \ln Y_1 }$, and $f_{\eta_{\theta,1} \mid \ln I_1,  \ln \theta_1,  \ln Y_0, \ln Y_1} = f_{\eta_{\theta,1}}$. Notice that $f_{Z_{\theta,1,m} \mid  \ln \theta_1 }$ and $f_{ \ln \theta_1 \mid  \ln Y_0}$ have already been estimated in the previous steps. Hence, the only remaining parameters are those of the measurement error equations for investment in period $1$ (i.e. $f_{Z_{I,1,m} \mid  \ln I_1 }$), the investment equation in period $1$ (i.e. $ f_{\ln I_1 \mid \ln \theta_1,  \ln Y_1}$), the density of the production function shock in period $1$ (i.e. $f_{\eta_{\theta,1}}$), as well as the measurement error equations for skills in period $2$ along with the production function (i.e. $f_{Z_{\theta,2,m}  \mid \ln \theta_1, \ln I_1, \eta_{\theta,1} } $).

This framework allows us to proceed iteratively for the remaining periods. First, we construct a synthetic dataset to estimate the density of $\ln \theta_2 \mid \ln Y_0, \ln Y_1$. We then estimate the relevant parameters for subsequent periods in a manner analogous to previous steps based on $f_{Z_{\theta,2}, Z_{I,2}, Z_{\theta,3} \mid \ln Y_0, \ln Y_1, \ln Y_3}$. This iterative procedure extents to general time periods $t$, involving $f_{Z_{\theta,t}, Z_{I,t}, Z_{\theta,t+1} \mid \ln Y_0, \ldots \ln Y_t}$ and the already estimated density $f_{\ln \theta_t \mid \ln Y_0, \ldots \ln Y_{t-1}}$. Consequently, the approach evaluates only low-dimensional integrals at each estimation step. While our estimator does not incorporate all available information from the complete joint likelihood, it balances efficiency with computational feasibility. 

\begin{remark}
    In the initial step, we estimate the conditional distribution $\ln \theta_0 \mid \ln Y_0$ using the skill measures $Z_{\theta_0}$ and income $Y_0$. Alternatively, one can also use both the skill and investment measures to estimate the joint distribution $\ln \theta_0, \ln I_0 \mid \ln Y_0$. The subsequent estimation steps would then rely on $f_{Z_{\theta,t}, Z_{I,t} ,Z_{\theta,t+1}, Z_{I,t+1} \mid \ln Y_0, \ldots, Y_{t+1}}$. 
    
    Our proposed estimation specification has two advantages. First, it allows for flexible estimation of the initial condition $\ln \theta_0 \mid \ln Y_0$, while remaining computationally tractable, as it does not, at the same time, estimate the measurement system of investment and the investment equation. The second advantage is more subtle: \citeN{Freyberger:24} shows that for the CES production function, the factor loading $\lambda_{I,0,1}$ is identified through the production function in period $t = 0$. Hence, we can identify $\lambda_{I,0,1}$ with the joint distribution of skill measures of the first two periods and the investment measure in period $0$. In contrast, it is not identified using only the skill and investment measures from period $0$.  
\end{remark}

\begin{remark}
For the likelihood function, we have to numerically approximate the integrals. We experimented with different methods in our simulations, including quadrature rules with a product grid, sparse grids, Monte Carlo integration, and Halton sequences. Halton sequences generally produced the most accurate estimators. Since we use a mixture normal for the skills in the initial period and normal distributions for the other unobserved variables, we transform uniform Halton draws to a grid on $\R$ by using the quantile function of the standard normal distribution.
\end{remark}

\subsection{Large sample distribution and bootstrap}
\label{s:bootstrap}
Our multiple step MLE is asymptotically normally distributed under standard regularity conditions. However, estimating the variance of the large sample distribution is challenging because it requires taking the estimation uncertainty of previously estimated parameters into account. For example, to calculate standard errors for $\hat{\delta}_2$, the estimated parameters of the production function in period $2$, we have to account for the fact that we estimated the skill distribution in period $1$, among others.  

Recall that we estimate all parameters of the model in $T+1$ steps, where each step estimates a subset of the parameters by maximizing likelihood functions. Let $\tau_t$ be the subset of the parameter vector estimated at time $t$ and let $\Tau_t$ be the parameter space. Let $\{W^{t}_i \}^n_{i=1}$ be the subset of the data used in step $t$. Let $W_i = \cup^{T+1}_{t=1} W^{t}_i $ and $\tau = (\tau_1, \ldots,\tau_{T+1})$. We denote the true value of $\tau$ and $\tau_{t}$ by $\tau_0$ and $\tau_{0,t}$, respectively. Notice that the parameter vectors are distinct in different time periods, but the samples overlaps. Also define  $\tau_{1:t} = (\tau_1, \ldots,\tau_t)$. Denote the likelihood in the first period by $l_1(W^{1}_i, \tau_1)$. For periods $t>1$, we use the notation $l_t(W^{t}_i, \tau_t \mid \tau_{t-1}, \ldots, \tau_{1} )$ to clarify that the likelihood depends on the parameters of the previous steps. Our estimator $\hat{\tau} = (\hat{\tau}_1, \ldots,\hat{\tau}_{T+1})$ is then
$$\hat{\tau}_1 = \argmin_{\tau_1 \in \Tau_1} \sum^n_{i=1} l_1(W^{1}_i, \tau_1)$$
and, for $t>1$,
$$\hat{\tau}_t = \argmin_{\tau_t \in \Tau_t} \sum^n_{i=1} l_t(W^{t}_i, \tau_t \mid \hat{\tau}_{t-1}, \ldots, \hat{\tau}_{1}).$$

It is easy to show that under standard regularity conditions
$$\sqrt{n}(\hat{\tau}_1 - \tau_{0,1}) = \left(  -\frac{1}{ n}\sum^n_{i=1} \frac{\partial^2}{ \partial \tau_1 \partial \tau_1'  }   l_1(W^{1}_i, \tau_{0,1} ) \right)^{-1}   \frac{1}{ \sqrt{n}}\sum^n_{i=1} \frac{\partial}{\partial \tau_1 } l_1(W^{1}_i, \tau_{0,1} ) + o_p(1)$$
and
\begin{align*}
    & \sqrt{n}(\hat{\tau}_t - \tau_{0,t}) \\
    & = \left( - \frac{1}{ n}\sum^n_{i=1} \frac{\partial^2}{ \partial \tau_t \partial \tau_t'  }   l_t(W^{t}_i, \tau_{0,t} \mid \tau_{0,1:t-1}) \right)^{-1}   \frac{1}{ \sqrt{n}}\sum^n_{i=1} \frac{\partial}{\partial \tau_t }  l_t(W^{t}_i, \tau_{0,t} \mid \hat{\tau}_{1:t-1}) + o_p(1) \\
    & = \left(  -\frac{1}{ n}\sum^n_{i=1} \frac{\partial^2}{ \partial \tau_t \partial \tau_t'  }   l_t(W^{t}_i, \tau_{0,t} \mid \tau_{0,1:t-1}) \right)^{-1}  \frac{1}{ \sqrt{n}}\sum^n_{i=1} \frac{\partial}{\partial \tau_t }  l_t(W^{t}_i, \tau_{0,t} \mid {\tau}_{0,1:t-1})  \\
       & \quad  -  \left(  \frac{1}{ n}\sum^n_{i=1} \frac{\partial^2}{ \partial \tau_t \partial \tau_t'  }   l_t(W^{t}_i, \tau_{0,t} \mid \tau_{0,1:t-1}) \right)^{-1}   \left( \frac{1}{ n} \sum^n_{i=1} \frac{\partial^2}{\partial \tau_t \partial \tau_{1:t-1}' }  l_t(W^{t}_i, \tau_{0,t} \mid {\tau}_{0,1:t-1}) \right) 
        \\
       & \qquad \quad \cdot \sqrt{n}( \hat{\tau}_{1:t-1} - {\tau}_{0,1:t-1} ) 
        + o_p(1). 
\end{align*}
Notice that if the parameters in previous periods were known, we would obtain
$$\sqrt{n}(\hat{\tau}_t - \tau_{0,t}) \stackrel{d}{\rightarrow} N\left( 0,\Gamma_t^{-1} V_t \Gamma_t^{-1}\right) $$
where 
$$\Gamma_t =  E\left[ \frac{\partial^2}{ \partial \tau_t \partial \tau_t'  } l(W^{t}_i, \tau_{0,t} \mid \tau_{0,t-1}, \ldots, \tau_{0,1}) \right] $$
and 
$$V_t = E\left[  \frac{\partial }{ \partial \tau_t    } l(W^{t}_i, \tau_{0,t} \mid \tau_{0,t-1}, \ldots, \tau_{0,1})  \frac{\partial }{ \partial \tau_t    } l(W^{t}_i, \tau_{0,t} \mid \tau_{0,t-1}, \ldots, \tau_{0,1})'   \right]$$
for $t>1$. In our setup, this asymptotic variance is incorrect because it ignores the estimation errors of $\hat{\tau}_{t-1}, \ldots, \hat{\tau}_{1}$, which is the second term in the expansion above. To account for those, we would have to calculate, among others, 
$$ \frac{\partial^2}{\partial \tau_t \partial \tau_{1:t-1} }  l_t(W^{t}_i, \tau_{t} \mid {\tau}_{1:t-1}')$$
for all $t>1$, which is very difficult because the likelihood is (partly) simulated and not available in closed form.

To avoid these calculations, we use a score bootstrap procedure inspired by \citeN{ABH:14}. This procedure does not require re-estimating any of the parts of the model, which would be computationally very costly. Let $\{W_i^*\}^n_{i=1}$ be a bootstrap sample, that we obtain by taking $n$ random draws from the original sample (with replacement). Let
$$\hat{\tau}_1^* = \hat{\tau}_1 - \left(  \frac1n \sum^n_{i=1} \frac{\partial^2}{ \partial \tau_1 \partial \tau_1'  } l_1(W^{1}_i, \hat{\tau}_1) \right)^{-1} \left( \frac{1}{n} \sum^n_{i=1}\frac{\partial }{ \partial \tau_1    } l_1(W^{1^*}_i, \hat{\tau}_1) 
 - \frac{1}{n} \sum^n_{i=1}\frac{\partial }{ \partial \tau_1    } l_1(W^{1}_i, \hat{\tau}_1) \right) $$
and
\begin{align*}
 \hat{\tau}_t^* &= \hat{\tau}_t - \left( \frac1n \sum^n_{i=1} \frac{\partial^2}{ \partial \tau_t \partial \tau_t'  } l(W^{t}_i, \hat{\tau}_t \mid \hat{\tau}_{1:t-1}) 
 \right)^{-1}  \\
 & \qquad \qquad \left( \frac{1}{n} \sum^n_{i=1}\frac{\partial }{ \partial \tau_t    }  l(W^{t^*}_i, \hat{\tau}_t \mid \hat{\tau}^*_{1:t-1}) 
 - \frac{1}{n} \sum^n_{i=1}\frac{\partial }{ \partial \tau_t    }  l(W^{t}_i, \hat{\tau}_t \mid \hat{\tau}_{1:t-1})  \right).  
\end{align*} 
Notice that, up to a negligible remainder term,
\begin{align*}
 &   \frac{1}{n} \sum^n_{i=1}\frac{\partial }{ \partial \tau_t    }  l(W^{t^*}_i, \hat{\tau}_t \mid \hat{\tau}^*_{1:t-1}) 
 - \frac{1}{n} \sum^n_{i=1}\frac{\partial }{ \partial \tau_t    }  l(W^{t}_i, \hat{\tau}_t \mid \hat{\tau}_{1:t-1})      \\
 &   =   \frac{1}{n} \sum^n_{i=1}\frac{\partial }{ \partial \tau_t    }  l(W^{t^*}_i, \hat{\tau}_t \mid \hat{\tau}_{1:t-1}) 
 - \frac{1}{n} \sum^n_{i=1}\frac{\partial }{ \partial \tau_t    }  l(W^{t}_i, \hat{\tau}_t \mid \hat{\tau}_{1:t-1})  \\
 & \quad + \left(\frac{1}{n} \sum^n_{i=1}\frac{\partial^2 }{ \partial \tau_t  \partial \tau_{1:t-1}'    }  l(W^{t^*}_i, \hat{\tau}_t \mid \hat{\tau}_{1:t-1}) \right) (\hat{\tau}^*_{1:t-1} - \hat{\tau}_{1:t-1} )
\end{align*} 
and we therefore correctly account for the estimation uncertainty of the pre-estimated parameters in periods $1, \ldots, t-1$. It now follows from an extension of the results of \shortciteN{ABH:14} that the asymptotic distribution of $\sqrt{n}(\hat{\tau} - \tau)$ can be consistently estimated by the distribution of $\sqrt{n}(\hat{\tau}^* - \hat{\tau})$.

A major advantage of this bootstrap procedure is that for each $t$, we only have to calculate derivatives of the likelihood with respect to $\tau_t$, but not with respect  to $\tau_s$ for $s<t$. These derivatives can either be calculated analytically or numerically.

\section{Monte Carlo simulations}
\label{s:simulations}

We start with a data generating process (DGP) as in \citeN{Freyberger:24}. This DGP is adapted from \shortciteN{AMN:19}, who state that it is designed to mimic their actual data. That is, we use 
$$
\theta_{t+1}= A_{t}\left(\gamma_{t} \theta_{t}^{\sigma_{t}}+\left(1-\gamma_{t}\right) I_t^{\sigma_{t}}\right)^{\frac{1}{\sigma_{t}} }\exp(\eta_{\theta,t})
$$
for $t=0,1$. While the coefficients in front of investment and skills sum to one, allowing for $A_t \neq 1$ is as general as not imposing this restriction. For our counterfactuals, it will be useful to also model the investment process and we let 
$$\ln I_t = \beta_{1t} \ln \theta_t + \beta_{2t} \ln Y_t  + \eta_{I,t}$$
where $Y_0 = Y_1 = Y$. To simulate data, we first draw $(\ln(\theta_{0}),\ln(Y))$ from a mixture of two normal distributions. In particular, the two components have means 
$$\mu_1 = (-4,-2)' \qquad \text{ and } \qquad \mu_2 = (6,3)'  $$
and variances
$$\Sigma_1 = \begin{pmatrix}
	0.620 & 0.035 \\ 0.035 & 0.056
\end{pmatrix}  \qquad \text{ and } \qquad
\Sigma_2 = \begin{pmatrix}
	0.83 & 0.17 \\ 0.17 & 1.28
\end{pmatrix}.$$
Notice that in this DGP of  \shortciteN{AMN:19}, the two normal distributions are very well separated. We will consider an alternative simulation setup with $\mu_1 = (3,1)$. We refer to the former results as ``original means'' and the latter as ``new means''. Given $(\ln(\theta_{0}),\ln(Y))$ and normally distributed $\eta_{\theta,t}$ and $\eta_{I,t}$,  we generate $I_0$, $\theta_1$,  $I_1$, and $\theta_2$ using the model. If $\ln I_t$ was equal to $\ln Y_t$, the setup would be exactly as in \shortciteN{AMN:19} with parameters as in their Table 9 and $\sigma_{0} = \sigma_{1} = -0.5$.\footnote{We use slightly different notation to be consistent with the notation above. Specifically, the periods are $t = 0,1,2$ instead of $t = 1,2,3$. We use $I_t$ instead of $X_t$ for the second latent variable and $\sigma_t$ instead of $\rho_t$ to denote the elasticity of substitution. The distribution of $(\ln(\theta_{0}),\ln(Y))$  is the same as the distribution of $(\ln(\theta_{0}),\ln(X))$ in \shortciteN{AMN:19}.}  We deviate slightly from their setting by using additional investment equations with $\beta_{1t} = 0.1$, $\beta_{2t} = 0.9$, and $\eta_{I,t}  \sim N(0,0.1^2)$. We simulate three measures for both $\theta_t$ and $I_t$, which have a factor structure. Following \shortciteN{AMN:19}, we set  $\mu_{\theta,t,m} = \mu_{I,t,m} = 0$ for all $m$ and $t$ and $\lambda_{\theta,t,1} = \lambda_{I,t,1} = 1$ for all $t$.

We implement both our estimator and that of \shortciteN{AMN:19} using a mixture of two normals, as in the original paper. This setup requires imposing scale and location restrictions to achieve point identification. For simplicity, we set $\mu_{\theta,t,m} = \mu_{I,t,m} = 0$ for all $t$ and $m$ and $\lambda_{\theta,t,1} = 1$ for all $t$. All other parameters in the measurement system, including all investment loadings, are free parameters to be estimated. We only report features that are invariant to scale and location restrictions - see \citeN{Freyberger:24}.

\subsection{Results with original means}

\begin{figure}[t!]
	\begin{center}
	\caption{Elasticities with original means}
	\label{f:elasticities_original}
\includegraphics[trim=3cm 4cm 3cm 0.5cm, width=14cm]
{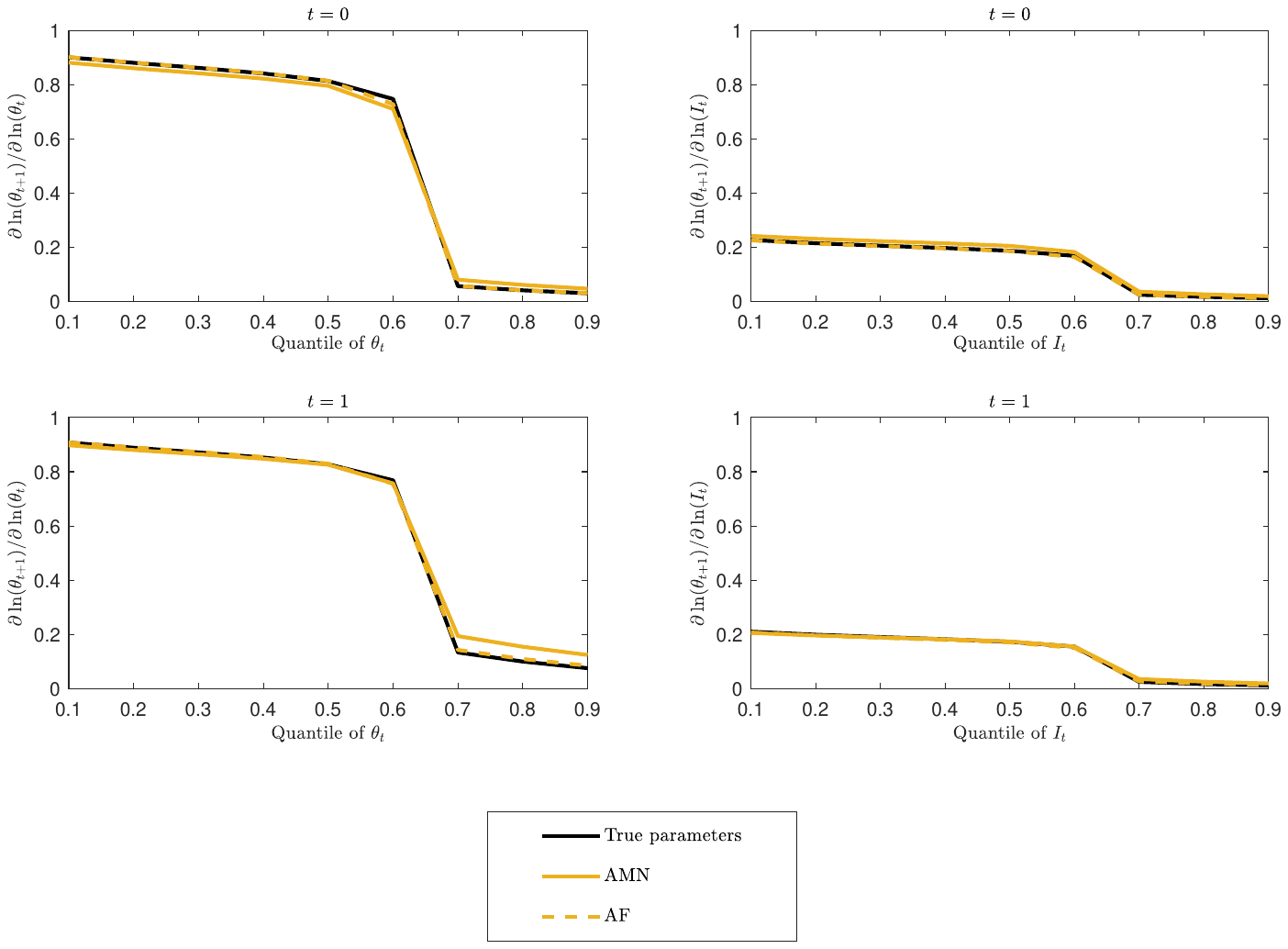}
	\begin{minipage}{0.9\textwidth}
		\vspace{5mm}
		{\footnotesize \begin{singlespace} Notes:  The figure shows estimated average elasticities of our estimator and that of  \shortciteN{AMN:19} based on the DGP of  \shortciteN{AMN:19}.
		\end{singlespace}} 
	\end{minipage}
		\end{center}
\end{figure}

This subsection contains the simulation results with $\mu_1 = (-4,-2)'$, as in \shortciteN{AMN:19}. Figure \ref{f:elasticities_original} shows average estimates of
$$\left.\frac{\partial \ln \theta_{t+1} }{\partial \ln \theta_{t} } \right|_{ \ln \theta_{t} = Q_{\alpha_1}(\ln \theta_{t}), \ln I_{t} = Q_{\alpha_2}(\ln I_{t})}      $$
for $\alpha_2 = 0.5$ and different values of $\alpha_1$ in the left panels and average estimates of 
$$\left.\frac{\partial \ln \theta_{t+1} }{\partial \ln I_{t} } \right|_{ \ln \theta_{t} = Q_{\alpha_1}(\ln \theta_{t}), \ln I_{t} = Q_{\alpha_2}(\ln I_{t})}      $$
for $\alpha_1 = 0.5$ and different values of $\alpha_2$ in the right panels. These results are based on $500$ Monte Carlo simulations and a sample size of $2000$. The figures show that both estimators perform well and have small biases. The main difference is in the lower left panel, where the estimator of \shortciteN{AMN:19} has a noticeable bias for large quantiles of $\theta_t$. These results are in line with \shortciteN{AMN:19}, who also find that the bias of their estimator is small.

Table \ref{tab:sim_elasticities_original} shows absolute biases and standard deviations of the estimators of
$$\left.\frac{\partial \ln \theta_{t+1} }{\partial \ln \theta_{t} } \right|_{ \ln \theta_{t} = Q_{\alpha_1}(\ln \theta_{t}), \ln I_{t} = Q_{\alpha_2}(\ln I_{t})}      $$
for $\alpha_2 = 0.5$ and two different sample sizes. The results are averaged over $\alpha_1 \in \{0.1, 0.2, \ldots, 0.9\}$ and labeled ``Skill elasticities''. In particular, for the bias, we calculate the absolute value of the bias for each $\alpha_1$ and report the average value. Hence, all biases are positive.  In addition, 

\clearpage

\begin{table}[b!]
	\caption{Simulation results for elasticities}
	\label{tab:sim_elasticities_original}
	\begin{center}
		\vspace{-1em}
		\begin{tabular}{lcccc}
			\hline \hline
			& Bias AMN & Std AMN & Bias AF & Std AF \\
			\hline
			&\multicolumn{4}{c}{ $n = 500$} \\
			Skill elasticity  for $t = 0$ 	& 0.0372 & 0.0690 & 0.0137 & 0.0517\\
			Skill elasticity  for $t = 1$ 		&  0.0398 & 0.0866 & 0.0199 & 0.0796\\
			Investment elasticity  for $t = 0$ 		& 0.0166 & 0.0484 & 0.0033 & 0.0302\\ 
			Investment elasticity  for $t = 1$		& 0.0078 & 0.0416 & 0.0052 & 0.0370  \\ 
			\hline
			& \multicolumn{4}{c}{ $n = 2000$} \\
			Skill elasticity  for $t = 0$ & 0.0216 & 0.0330 & 0.0027 & 0.0232\\ 
			Skill elasticity  for $t = 1$ 	& 0.0229 & 0.0400 & 0.0053 & 0.0385\\
			Investment elasticity  for $t = 0$ 		&0.0140 & 0.0238 & 0.0012 & 0.0143\\
			Investment elasticity  for $t = 1$		&0.0042 & 0.0191 & 0.0027 & 0.0186\\
			\hline
			\hline
		\end{tabular}
        	\begin{minipage}{0.83\textwidth}
		\vspace{5mm}
		{\footnotesize \begin{singlespace} Notes:  The table shows absolute biases and standard deviations of average  elasticities of our estimator and that of  \shortciteN{AMN:19} based on the DGP of  \shortciteN{AMN:19}.
		\end{singlespace}} 
	\end{minipage}
	\end{center}
\end{table}

 \clearpage

\noindent the table contains absolute biases and standard deviations  of the estimators of
$$\left.\frac{\partial \ln \theta_{t+1} }{\partial \ln I_{t} } \right|_{ \ln \theta_{t} = Q_{\alpha_1}(\ln \theta_{t}), \ln I_{t} = Q_{\alpha_2}(\ln I_{t})}      $$
for $\alpha_1 = 0.5$ and averaged over $\alpha_2 \in \{0.1, 0.2, \ldots, 0.9\}$, labeled ``Investment elasticities''. These results confirm that both estimators generally have small biases and also show that the standard deviations are comparable. Even though the biases are all small, for the larger sample size, our estimator still has substantially smaller biases.

\begin{figure}[t!]
	\begin{center}
	\caption{Quantile effects with original means}
	\label{f:quantiles_original}
    	\includegraphics[trim=3cm 4cm 3cm 0cm, width=14.0cm]{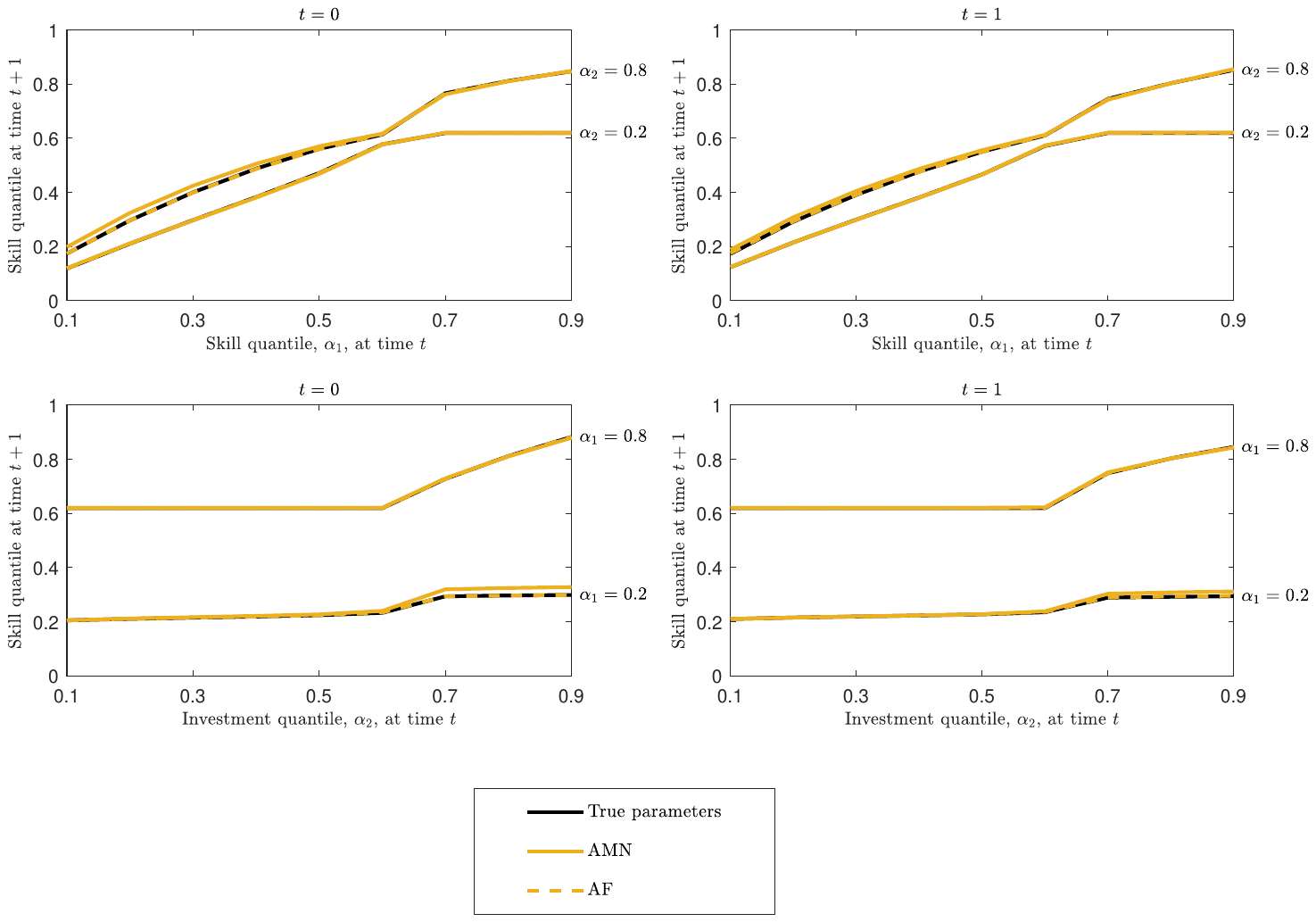}
		\begin{minipage}{0.9\textwidth}
		\vspace{5mm}
		{\footnotesize \begin{singlespace} Notes:  The figure shows estimated average quantile effects of our estimator and that of  \shortciteN{AMN:19} based on the DGP of  \shortciteN{AMN:19}.
		\end{singlespace}} 
	\end{minipage}
	\end{center}
\end{figure}

Figure \ref{f:quantiles_original} shows average estimates of
$$F_{\theta_{t+1}}\left( A_t\left(\gamma_{t} Q_{\alpha_1}(\theta_t)^{\sigma_t}   +  (1-\gamma_{t} )Q_{\alpha_2}(I_t)^{\sigma_t} \right)^{\frac{1}{\sigma_t}}    \right) $$
for different values of $\alpha_1$ and $\alpha_2$. These results show the effects of changes in skills and investments at time $t$ on the rank in the skill distribution at time $t+1$. Again, both estimators only have small biases.

Table \ref{tab:sim_quantile_original} displays the corresponding absolute biases and standard deviations of estimators (times 10) for $\alpha_2 = 0.5$, averaged over $\alpha_1 \in \{0.1, 0.2, \ldots, 0.9\}$ (labeled ``Skill effect'') and $\alpha_1 = 0.5$, averaged over $\alpha_2 \in \{0.1, 0.2, \ldots, 0.9\}$ (labeled ``Investment effect''). Again, these results illustrate that the biases are small and that the standard deviations are comparable.

\begin{figure}[h]
	\begin{center}
	\caption{Counterfactual distributions  with original means }
	\label{f:distributions_original}
	\includegraphics[trim=2cm 0.5cm 3cm 0cm, width=15cm]
{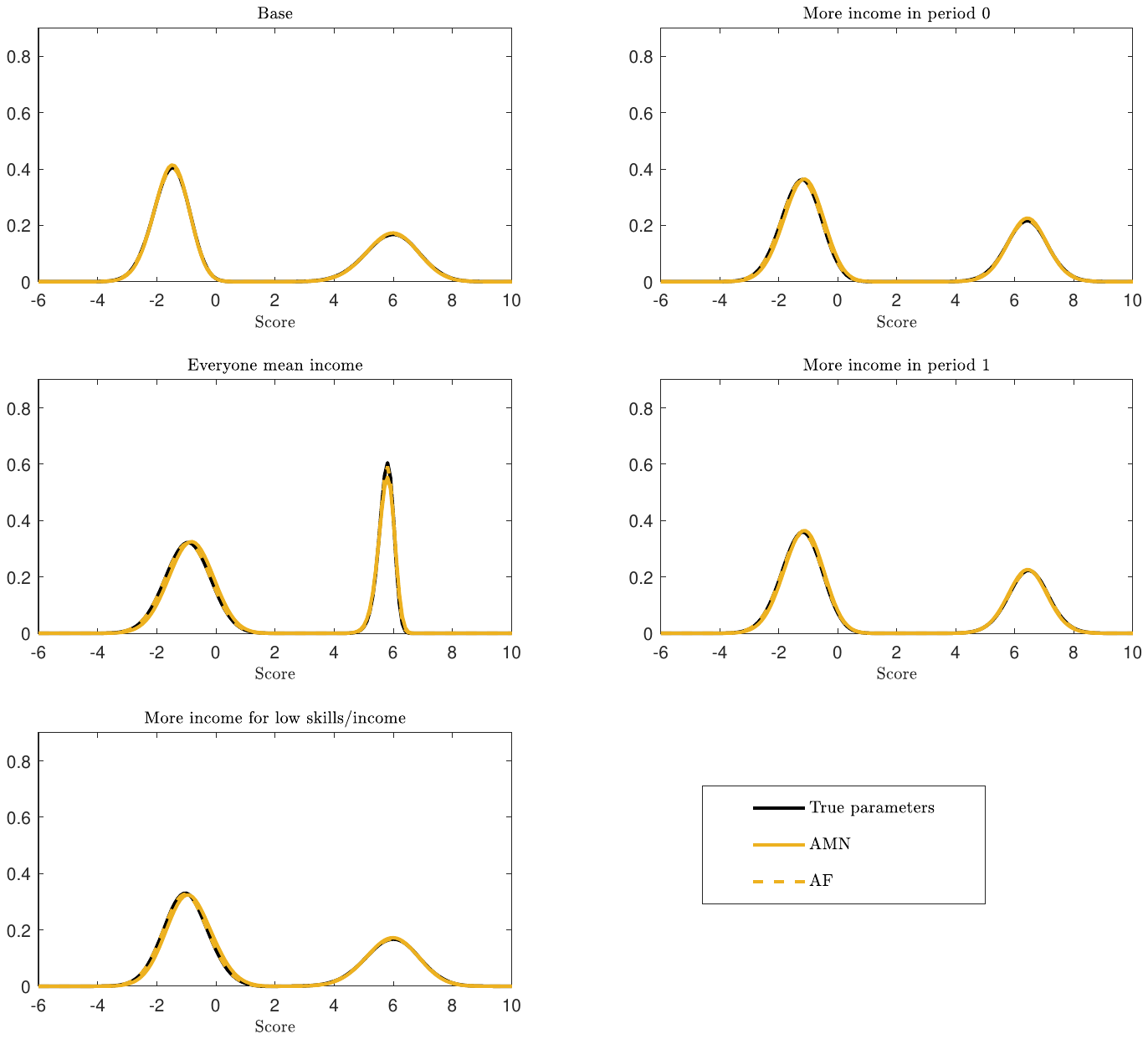}
    
    	\begin{minipage}{0.8\textwidth}
		\vspace{5mm}
		{\footnotesize \begin{singlespace} Notes:  The figure shows estimated average counterfactual distributions of one of the measures. It compares  our estimator and that of  \shortciteN{AMN:19} based on the DGP of  \shortciteN{AMN:19}.
		\end{singlespace}} 
	\end{minipage}
		\end{center}
\end{figure}

Next, we analyze how exogenous income changes affect the skill distribution. The results of this counterfactual for the estimator of \shortciteN{AMN:19} have also been reported by \citeN{Freyberger:24}. Following that paper, we first take draws from the estimated joint distribution of income and skills in period $0$  and consider four counterfactual marginal income distributions. First, we increase everyone's income by two standard deviations in period 0. Second, we increase everyone's income by two standard deviations in period 1. Third, we set income to the median for everyone in both periods. Fourth, we increase income by two standard deviations in both periods, but only if the initial skill and income quantiles are below $0.5$. We set all unobservables to their median values. Figure \ref{f:distributions_original} shows the averages of these estimated counterfactual test score distributions. Both estimator match the  baseline and the counterfactual distributions well.

\clearpage

\begin{table}[b!]
	\caption{Simulation results for quantile effects}
	\label{tab:sim_quantile_original}
	\begin{center}
		\vspace{-1em}
		\begin{tabular}{lcccc}
			\hline \hline
			& Bias AMN & Std AMN & Bias AF & Std AF \\
			\hline
			&\multicolumn{4}{c}{ $n = 500$} \\
			Skill effect  for $t = 0$ 	&   0.0159 & 0.1095 & 0.0089 & 0.0997\\
			Skill effect  for $t = 1$ 		&0.0092 & 0.1049 & 0.0108 & 0.1049\\
			Investment effect  for $t = 0$ 		& 0.0567 & 0.1474 & 0.0107 & 0.1046\\
			Investment effect  for $t = 1$		& 0.0319 & 0.1369 & 0.0099 & 0.1268  \\ 
			\hline
			& \multicolumn{4}{c}{ $n = 2000$} \\
			Skill effect  for $t = 0$ &0.0149 & 0.0562 & 0.0032 & 0.0475\\
			Skill effect  for $t = 1$ 	&0.0091 & 0.0545 & 0.0034 & 0.0516\\
			Investment effect  for $t = 0$ 	&	0.0480 & 0.0735 & 0.0059 & 0.0498\\
			Investment effect  for $t = 1$		& 0.0218 & 0.0691 & 0.0060 & 0.0680 \\
			\hline
			\hline
		\end{tabular}
         	\begin{minipage}{0.80\textwidth}
		\vspace{5mm}
		{\footnotesize \begin{singlespace} Notes:  The table shows biases and standard deviations of quantile effects of our estimator and that of  \shortciteN{AMN:19} based on the DGP of \shortciteN{AMN:19}.
		\end{singlespace}} 
	\end{minipage}
	\end{center}
\end{table}  

\clearpage

\clearpage 

\subsection{Results with new means}

\begin{figure}[t!]
	\begin{center}
	\caption{Elasticities with new means}
	\label{f:elasticities_new_means_3_1}
    	\includegraphics[trim=3cm 4cm 3cm 0cm, width=14.5cm]{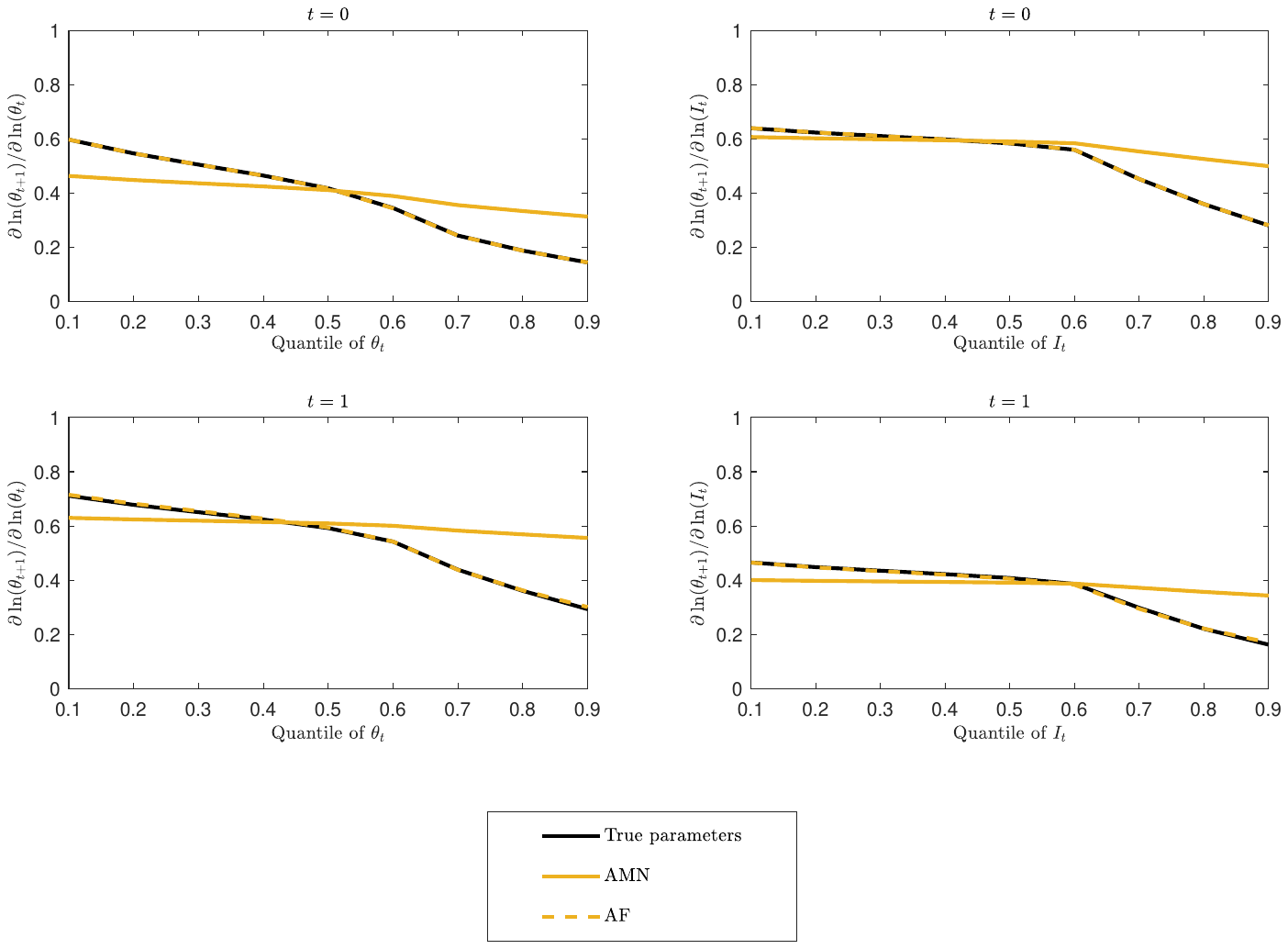}
        	\begin{minipage}{0.9\textwidth}
		\vspace{5mm}
		{\footnotesize \begin{singlespace} Notes: As Figure \ref{f:elasticities_original}, but with $\mu_1 = (3,1)'$.
		\end{singlespace}} 
	\end{minipage}
		\end{center}
\end{figure}

As can be seen from Figure \ref{f:distributions_original}, the two modes of the implied test score distributions are very well separated. The distribution might therefore not be a good description of common data sets. We now repeat the simulation exercises but use $\mu_1 = (3,1)'$ instead. All other parameters are unchanged. 
Figure \ref{f:elasticities_new_means_3_1} shows average estimates of
$$\left.\frac{\partial \ln \theta_{t+1} }{\partial \ln \theta_{t} } \right|_{ \ln \theta_{t} = Q_{\alpha_1}(\ln \theta_{t}), \ln I_{t} = Q_{\alpha_2}(\ln I_{t})}  \quad  \text{ and } \quad  \left.\frac{\partial \ln \theta_{t+1} }{\partial \ln I_{t} } \right|_{ \ln \theta_{t} = Q_{\alpha_1}(\ln \theta_{t}), \ln I_{t} = Q_{\alpha_2}(\ln I_{t})}      $$
analogous to Figure \ref{f:elasticities_original}. In this setup, the estimator of \shortciteN{AMN:19} based an a mixture of 2 normals is significantly biased, as opposed to our estimator.

\begin{figure}[t!]
	\begin{center}
	\caption{Density of centered elasticities with new means}
	\label{f:elasticities_density_new_means_3_1}
    	\includegraphics[trim=3cm 4cm 3cm 0cm, width=14.5cm]{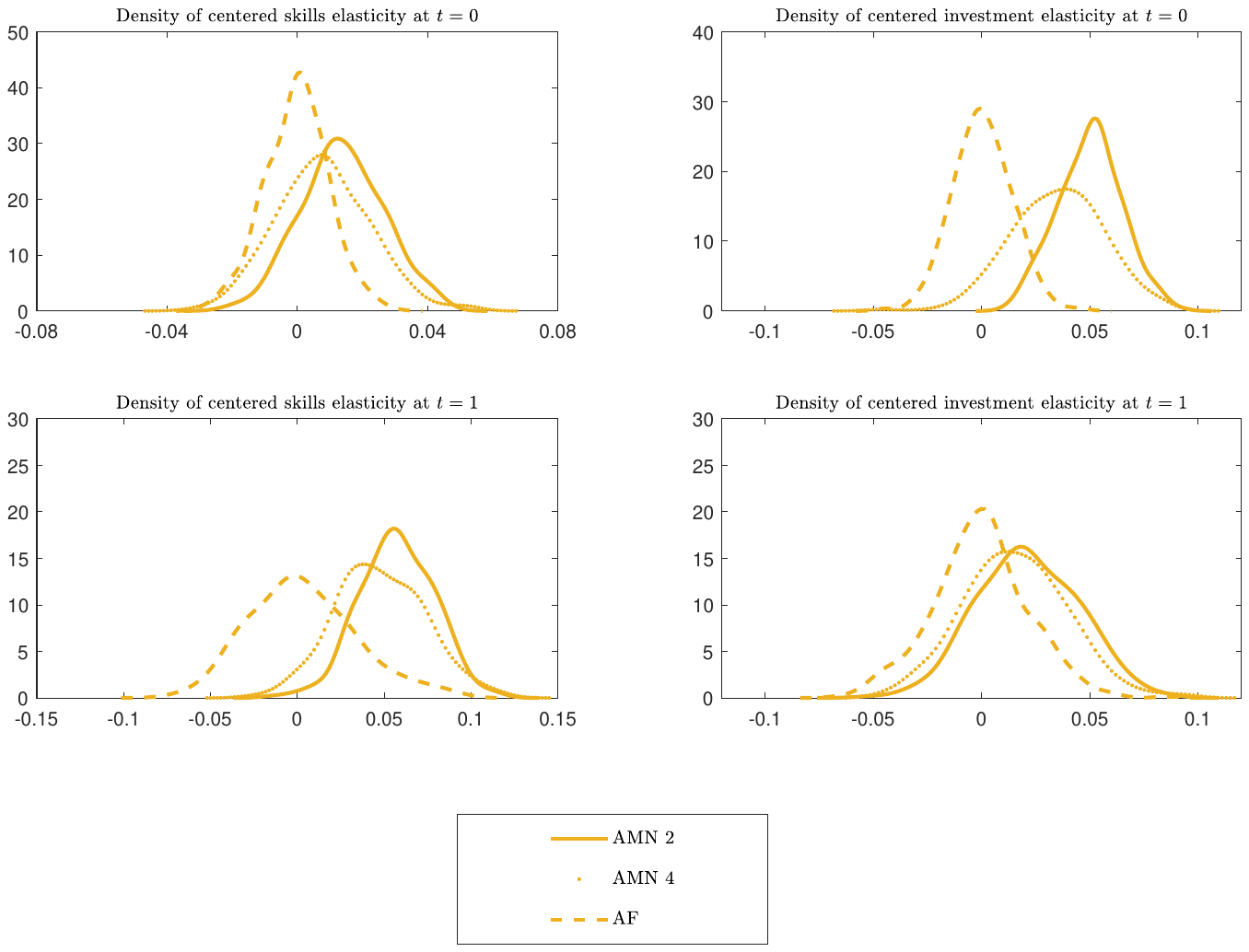}
        	\begin{minipage}{0.9\textwidth}
		\vspace{5mm}
		{\footnotesize \begin{singlespace} Notes: The figure shows kernel density estimators of the centered estimates of the average elasticities. The results based on \shortciteN{AMN:19} with mixtures of 2 and 4 normals are denoted by AMN 2 and AMN 4, respectively.
		\end{singlespace}} 
	\end{minipage}
		\end{center}
\end{figure}

Table \ref{tab:sim_elasticities_new_means_3_1} shows the corresponding biases and standard deviations. For the estimator of \shortciteN{AMN:19}, we use both a mixture of two normals as well as a mixture of four normals. Using a larger number of components decreases the bias slightly, but it comes at the expense of larger standard errors. Moreover, the estimator can be numerically unstable. In particular, when $n=2000$, in 1\% of the simulated data sets the EM algorithm failed to converge. The reason is that the means and variances of the mixture components are only weakly identified if the number of mixture components is too large and the estimated components are not sufficiently well separated. Figure \ref{f:elasticities_density_new_means_3_1} shows the density of the estimates corresponding to Table \ref{tab:sim_elasticities_new_means_3_1} centered at the true value. For the skill and investment elasticities, we average over different values of $\alpha_2$ and $\alpha_1$, respectively.  Since the bias of our estimator is very small, the distributions are centered at $0$. The biases of the estimated elasticities based on  \shortciteN{AMN:19} are negative for small quantiles and positive for large quantiles (see Figure \ref{f:elasticities_new_means_3_1}). Hence, the average biases is smaller than the average absolute biases reported in Table \ref{tab:sim_elasticities_new_means_3_1}, but they are still noticably different from 0.

\begin{figure}[t!]
	\begin{center}
	\caption{Quantile effects with new means}
	\label{f:quantiles_new_means_3_1}
	\includegraphics[trim=3cm 4cm 3cm 0cm, width=14.5cm]{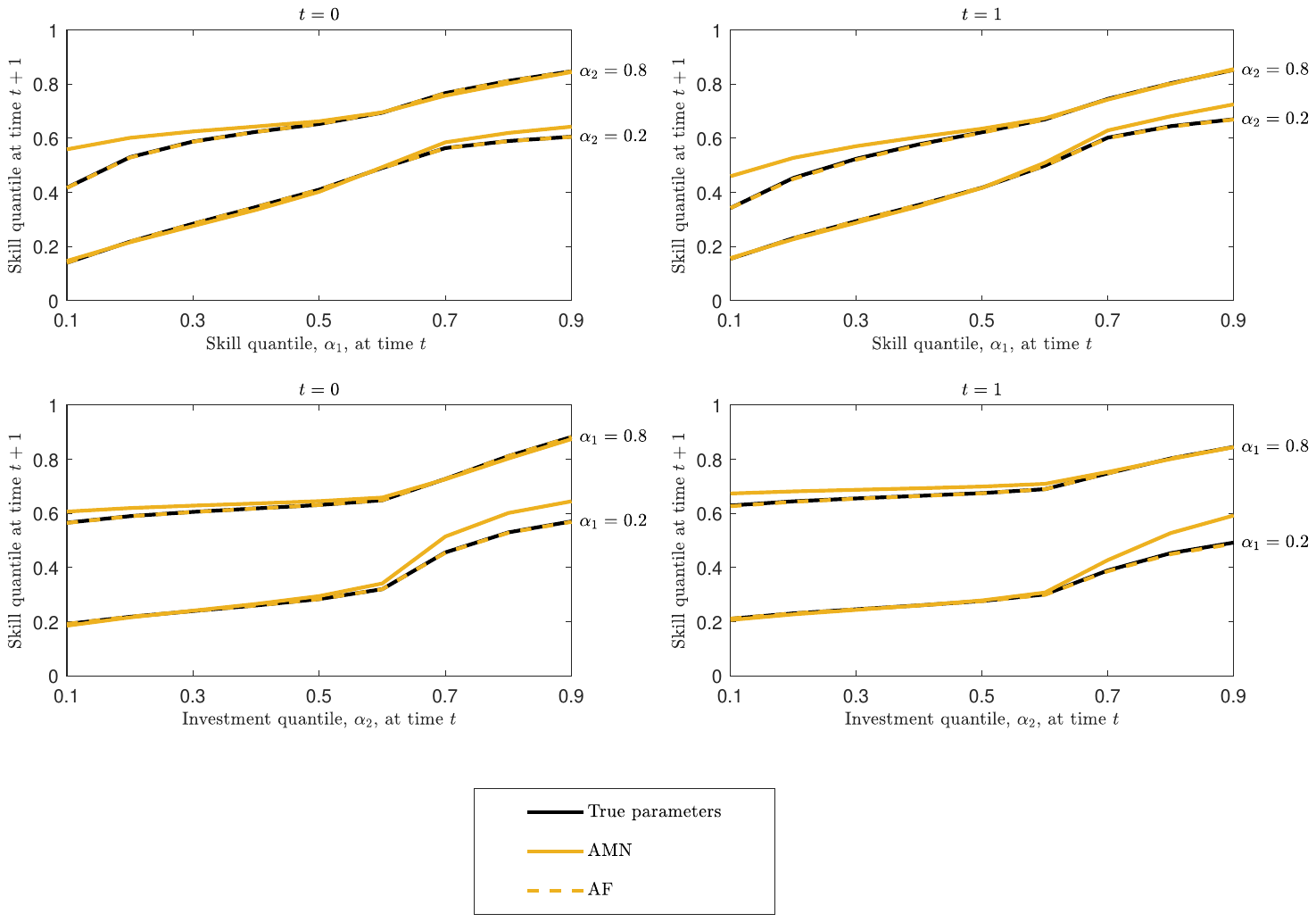}
            	\begin{minipage}{0.9\textwidth}
		\vspace{5mm}
		{\footnotesize \begin{singlespace} Notes: As Figure \ref{f:quantiles_original}, but with $\mu_1 = (3,1)'$.
		\end{singlespace}} 
	\end{minipage}
		\end{center}
\end{figure}

\begin{figure}[t!]
	\begin{center}
	\caption{Density of centered quantile effects with new means}
	\label{f:qunatile_density_new_means_3_1}
    	\includegraphics[trim=3cm 4cm 3cm 0cm, width=14.5cm]{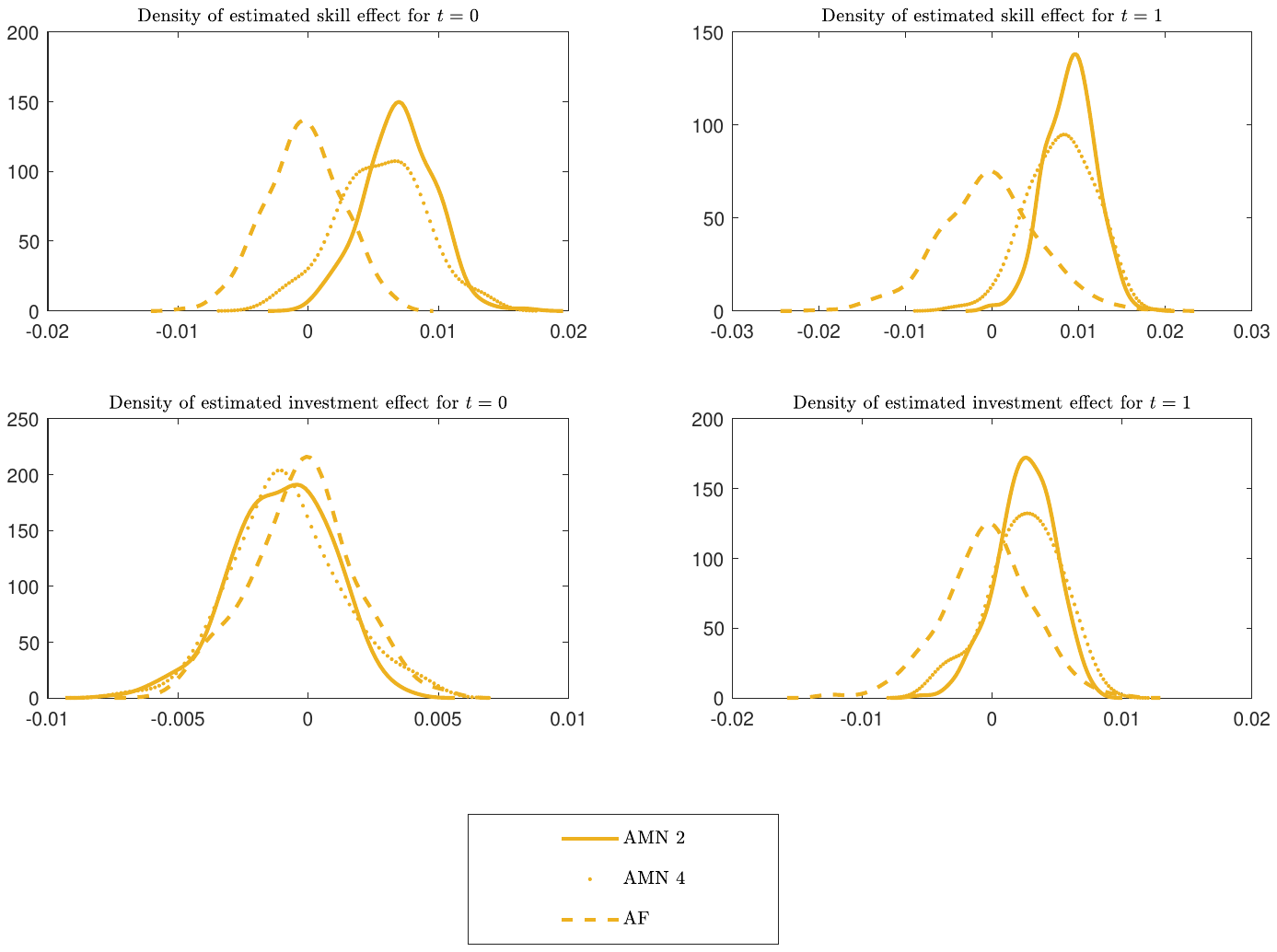}
        	\begin{minipage}{0.9\textwidth}
		\vspace{5mm}
		{\footnotesize \begin{singlespace} Notes: The figure shows kernel density estimators of the centered quantile effects of the average elasticities. The results based on \shortciteN{AMN:19} with mixtures of 2 and 4 normals are denoted by AMN 2 and AMN 4, respectively.
		\end{singlespace}} 
	\end{minipage}
		\end{center}
\end{figure}

Figure \ref{f:quantiles_new_means_3_1} and Table \ref{tab:sim_quantile_new_means_3_1} are analogous to Figure \ref{f:quantiles_original} and Table \ref{tab:sim_quantile_original}, respectively. Similar to the elasticities, the estimator of \shortciteN{AMN:19} has much large biases than our estimator.  
Figure \ref{f:qunatile_density_new_means_3_1} shows the density of the estimates corresponding to Table \ref{tab:sim_quantile_new_means_3_1} centered at the true value. While our estimator has a larger standard deviation in some cases, the estimates are centered at the true value. Moreover, the standard deviations are generally quite small.

\clearpage

\begin{table}[b!]
	\caption{Simulation results for elasticities}
	\label{tab:sim_elasticities_new_means_3_1}
	\begin{center}
		\vspace{-1em}
		\begin{tabular}{lcccccc}
			\hline \hline
			&\multicolumn{2}{c}{AMN  $2$ mixtures}  &\multicolumn{2}{c}{AMN  $4$ mixtures} && \\
			& Bias   & Std  & Bias   & Std   & Bias AF & Std AF \\
			\hline
			&\multicolumn{6}{c}{ $n = 500$} \\			
			Skill elasticity  for $t = 0$ 	&0.0920 & 0.0363 & 0.0651  &  0.0469 & 0.0017 & 0.0375\\
			Skill elasticity  for $t = 1$ 		&  0.0946 & 0.0685 & 0.0727 &   0.0790 & 0.0071 & 0.1021\\
			Investment elasticity  for $t = 0$ 		&0.0659 & 0.0327 & 0.0477  &  0.0420 & 0.0014 & 0.0353\\ 
			Investment elasticity  for $t = 1$	& 0.0663 & 0.0582 & 0.0522  &  0.0645  & 0.0066 & 0.0759  \\
			\hline
			& \multicolumn{6}{c}{ $n = 2000$} \\
			Skill elasticity  for $t = 0$ &   0.0911 & 0.0179 & 0.0619  &  0.0277 & 0.0003 & 0.0187\\
			Skill elasticity  for $t = 1$ &	0.0963 & 0.0323 &  0.0778  &  0.0405  & 0.0032 & 0.0569\\
			Investment elasticity  for $t = 0$ 	&  0.0652 & 0.0166 & 0.0453 &   0.0254  & 0.0004 & 0.0177\\  
			Investment elasticity  for $t = 1$	& 0.0659 & 0.0287 & 0.0531 &   0.0335  & 0.0025 & 0.0405 \\
			\hline
			\hline
		\end{tabular}
        \begin{minipage}{0.95\textwidth}
		\vspace{5mm}
		{\footnotesize \begin{singlespace} Notes: As Table \ref{tab:sim_elasticities_original}, but with $\mu_1 = (3,1)'$ and two different number of mixtures.
		\end{singlespace}} 
	\end{minipage}
	\end{center}
\end{table}

\clearpage

\clearpage

\begin{table}[b!]
	\caption{Simulation results for quantile effects}
	\label{tab:sim_quantile_new_means_3_1}
	\begin{center}
		\vspace{-1em}
		\begin{tabular}{lccc cc c}
			\hline \hline
			&\multicolumn{2}{c}{AMN  $2$ mixtures}  &\multicolumn{2}{c}{AMN  $4$ mixtures} && \\
			& Bias   & Std  & Bias   & Std   & Bias AF & Std AF \\	
            \hline
			&\multicolumn{6}{c}{ $n = 500$} \\
			Skill effect  for $t = 0$  &	0.1104 & 0.1029 & 0.0879  &  0.1127  & 0.0067 & 0.0986\\
			Skill effect  for $t = 1$ 		&  0.1203 & 0.1128 & 0.0907  &  0.1200 & 0.0107 & 0.1509\\
			Investment effect  for $t = 0$ 		&  0.0991 & 0.1043 & 0.0854  &  0.1013 & 0.0075 & 0.0814\\
			Investment effect  for $t = 1$		& 0.0648 & 0.1258 &  0.0410  &  0.1311 & 0.0147 & 0.1313\\
            \hline
			& \multicolumn{6}{c}{ $n = 2000$} \\
			Skill effect  for $t = 0$ &  0.1112 & 0.0504 & 0.0831  &  0.0597  & 0.0037 & 0.0476\\ 
			Skill effect  for $t = 1$ &	0.1256 & 0.0533 & 0.1040  &  0.0639 & 0.0052 & 0.0810\\
			Investment effect  for $t = 0$ 	&  0.0958 & 0.0463 & 0.0712 &    0.0578 & 0.0036 & 0.0407\\ 
			Investment effect  for $t = 1$	& 0.0699 & 0.0575 &  0.0570   & 0.0665  & 0.0077 & 0.0703\\
			\hline
			\hline
		\end{tabular}
           \begin{minipage}{0.95\textwidth}
		\vspace{5mm}
		{\footnotesize \begin{singlespace} Notes: As Table \ref{tab:sim_quantile_original}, but with $\mu_1 = (3,1)'$ and two different number of mixtures.
		\end{singlespace}} 
	\end{minipage}
	\end{center}
\end{table}

\clearpage

\clearpage

\begin{figure}[h]
	\begin{center}
	\caption{Counterfactual distributions  with new means }
	\label{f:distributions_new_means_3_1}
	\includegraphics[trim=2cm 0.5cm 3cm 0cm, width=15cm]
{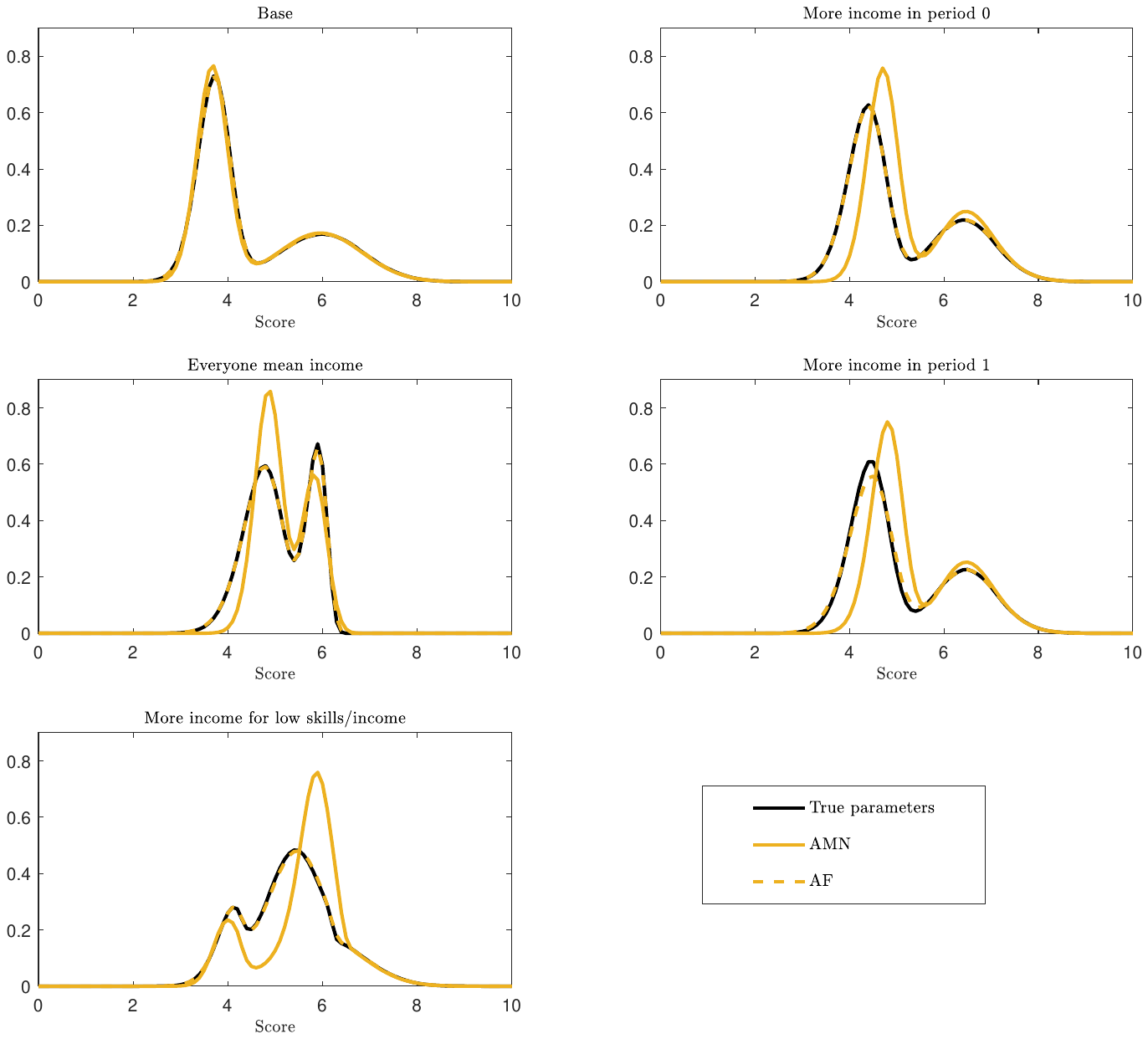}    
    	\begin{minipage}{0.8\textwidth}
		\vspace{5mm}
		{\footnotesize \begin{singlespace} Notes: As Figure \ref{f:distributions_original}, but with $\mu_1 = (3,1)'$. 
		\end{singlespace}} 
	\end{minipage}
		\end{center}
\end{figure}

Figure \ref{f:distributions_new_means_3_1} shows the averages of the estimated counterfactual test score distributions under different counterfactual income distributions. Interestingly, the estimator of \shortciteN{AMN:19} still matches the baseline distribution well, but yields biased counterfactual distributions. Specifically, it overestimates the effects of income changes, especially for low quantiles of the score distribution.

\begin{figure}[t!]
	\begin{center}
	\caption{Quantile paths with new means}
	\label{f:quantile_paths_simulations_ces_3_1}
	\includegraphics[trim=4cm 8.5cm 4cm 4cm, width=15cm]{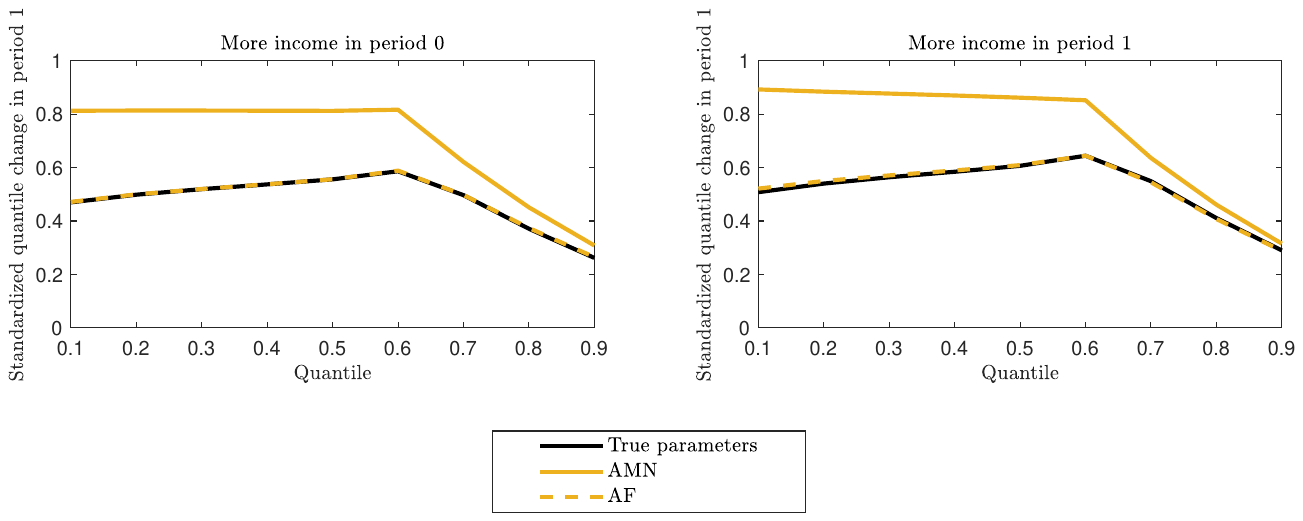}
    	\begin{minipage}{0.95\textwidth}
		\vspace{5mm}
		{\footnotesize \begin{singlespace} Notes:  The figure shows the standardized change in the quantile of the skill distributions
        for transfers in $t = 0$ and $t = 1$ based on the CES production function with $\mu_1 = (3,1)'$.
		\end{singlespace}} 
	\end{minipage}
		\end{center}
\end{figure}

We now consider the timing of investment - either in period 0 or in period 1 - in more detail. Figure \ref{f:quantile_paths_simulations_ces_3_1}  shows the standardized changes in the quantiles of the skill distributions for earlier and later transfers. For each quantile, the y-axis shows the difference of quantiles of the counterfactual and the baseline distribution divided by the standard deviation of the baseline distribution. For example, a value of $0.5$ for $\alpha = 0.1$ means that the income transfer increases the 0.1-quantile by $0.5$ standard deviations. Again the estimator of \citeN{AMN:19} is substantially biased. For example, for $\alpha = 0.1$ the true effects are $0.47$ and $0.51$ for transfers in period 0 and 1, but the (average) estimated effects are $0.81$ and $0.89$, respectively.

\vspace{-0.5cm}

\subsection{Results for trans-log production function}

We next return to the original DGP of \citeN{AMN:19}, but consider a trans-log production function that is the best approximation of the previous CES production function. In this case, we can also use the estimator of \citeN{AW:22}. 

As before, we consider elasticities (in Table \ref{tab:sim_elasticities_new_means_translog}) and quantile effects (in Table \ref{tab:sim_quantile_new_means_translog}). The results highlight that both our estimator and that of \citeN{AW:22} have small biases, but our estimator has a smaller standard deviation. We should note that \citeN{AW:22} do not use all valid moment conditions implied by the model. It may be interesting to investigate how all of these moments conditions can be used efficiently. An advantage of a likelihood-based approach is that it uses most/all of the available information. 

\begin{figure}[b!]
	\begin{center}
	\caption{Coverage probabilities and average lengths}
	\label{f:coverage_length}
	\includegraphics[trim=3cm 4.5cm 3cm 3.4cm, width=14.4cm]{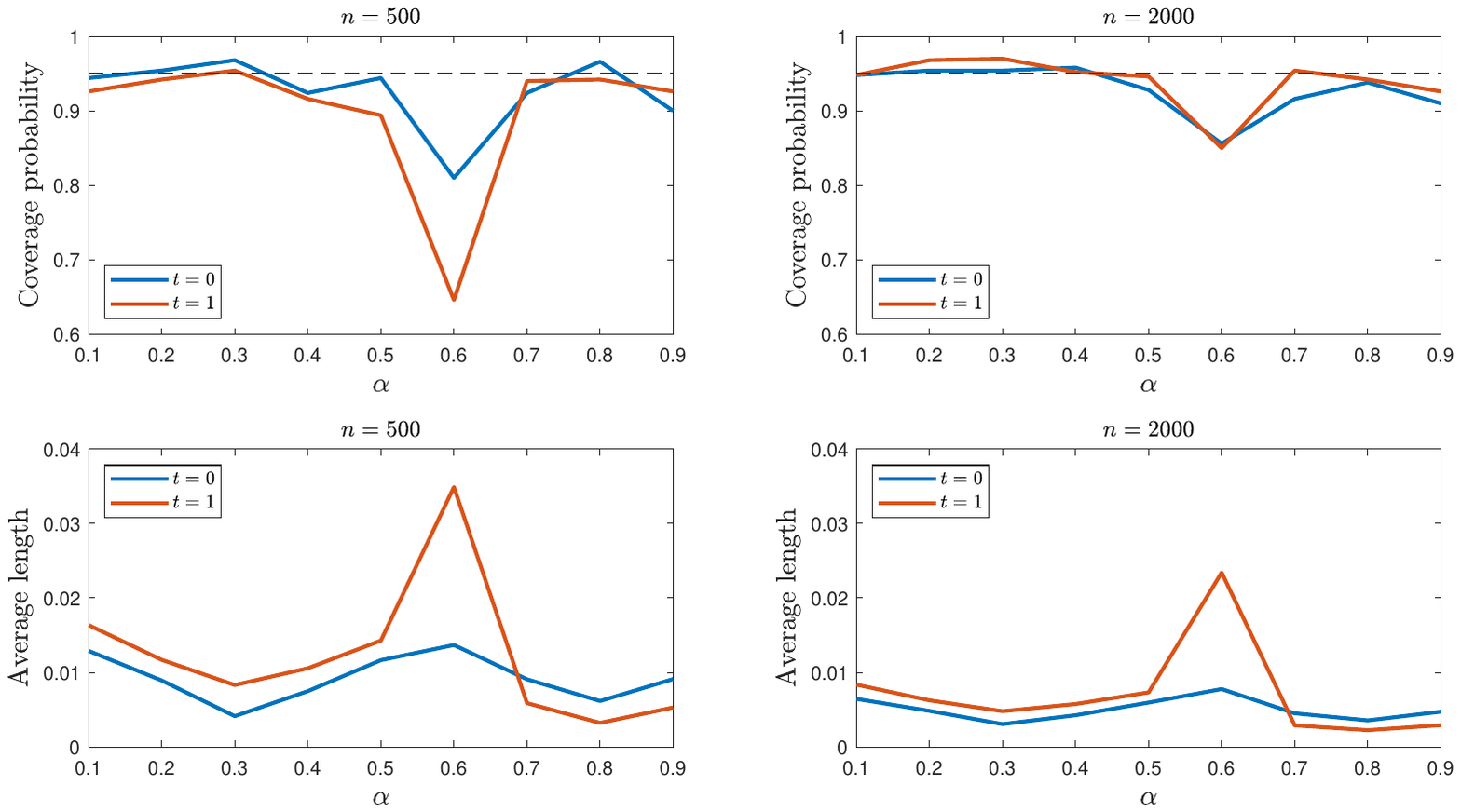}
    	\begin{minipage}{0.95\textwidth}
		\vspace{10mm}
		{\footnotesize \begin{singlespace} Notes: The figure shows coverage probabilities and average lengths of confidence intervals for $F_{\theta_{t+1}}\left( a_t + \gamma_{1t} Q_{\alpha}(\ln(\theta_t))  + \gamma_{2t} Q_{\alpha}(\ln(I_t))+ \gamma_{3t} Q_{\alpha}(\ln(\theta_t))Q_{\alpha}(\ln(I_t))  \right)$ for different values of $\alpha$. 
		\end{singlespace}} 
	\end{minipage}
		\end{center}
\end{figure}

Tables  \ref{tab:sim_elasticities_new_means_translog} and \ref{tab:sim_quantile_new_means_translog} also display coverage probabilities of confidence intervals for our estimator based on the bootstrap. Recall that the estimates of the quantile effects are either evaluated at the median of investment or the median of skills and at different quantiles of the other latent variable. Similar to the the biases and standard deviations, the coverage probabilities are averaged across these quantiles. Most of the coverage probabilities are close to the nominal level of $0.95$, especially for $n = 2000$. The slight undercoverage, even for $n=2000$, is due to quantile $0.6$. Here, the cdf is extremely steep since the two mixtures are very well separated, leading to visible kinks in the elasticities and the quantile functions (see Figures \ref{f:elasticities_original} and \ref{f:quantiles_original}), which makes estimation and inference particularly difficult (for any method). To illustrate this point, Figure \ref{f:coverage_length} displays coverage probabilities and average lengths of confidence intervals for $F_{\theta_{t+1}}\left( a_t + \gamma_{1t} Q_{\alpha}(\ln(\theta_t))  + \gamma_{2t} Q_{\alpha}(\ln(I_t))+ \gamma_{3t} Q_{\alpha}(\ln(\theta_t))Q_{\alpha}(\ln(I_t))  \right)$ for different values of $\alpha$. For $n= 2000$, the coverage probabilities are close to $0.95$ for all $\alpha \neq 0.6$.  

Recall that our bootstrap is computationally attractive as it does not require any numerical optimization. On a desktop computer (Intel(R) Core(TM) i9-10900 CPU @ 2.80GHz, 2801 Mhz, 10 Core(s), 64 GB RAM), the marginal costs of obtaining a bootstrap sample are around $2.5$ seconds when $n=500$ and $10$ seconds when $n=2000$.\\

\begin{table}[h!]
	\caption{Simulation results for elasticities}
	\label{tab:sim_elasticities_new_means_translog}
	\begin{center}
    %\begin{small}    
    	\vspace{-1em}
		\begin{tabular}{lccccccc}
			\hline \hline
			&  Bias  AMN  &  Std AMN &  Bias AW  & Std  AW & Bias AF & Std AF & Cov AF \\	
            \hline
           & \multicolumn{7}{c}{ $n = 500$} \\
			Skill elast.  $t = 0$ &   0.0004 & 0.0175  & 0.0025 & 0.0279 & 0.0012 & 0.0164 & 0.934\\ 
			Skill elast.  $t = 1$  & 0.0012 & 0.0304  & 0.0035 & 0.0342 & 0.0013 & 0.0300 & 0.910\\
			Inv. elast.  $t = 0$ &  0.0062 & 0.0699  & 0.0081 & 0.0953 &  0.0005 & 0.0555 & 0.932 \\ 
			Inv. elast.  $t = 1$  &	 0.0049 & 0.0852 & 0.0049 & 0.0852 &  0.0007 & 0.0840  &  0.886 \\
			\hline
			& \multicolumn{7}{c}{ $n = 2000$} \\
			Skill elast.  $t = 0$ & 0.0005 & 0.0096  & 0.0001 & 0.0134 & 0.0001 & 0.0077 & 0.950\\
			Skill elast.  $t = 1$  &   0.0011 & 0.0147  & 0.0016 & 0.0177 & 0.0010 & 0.0149 & 0.924\\ 
			Inv. elast.  $t = 0$ & 0.0038 & 0.0337 & 0.0006 & 0.0478 & 0.0005 & 0.0287 & 0.926 \\
			Inv. elast.   $t = 1$  &0.0029 & 0.0424 & 0.0029 & 0.0424 & 0.0040 & 0.0421  & 0.918  \\
			\hline
			\hline
		\end{tabular}
        %\end{small}
        \begin{minipage}{0.99\textwidth}
		\vspace{5mm}
		{\footnotesize \begin{singlespace} Notes:  The table shows biases and standard deviations of average elasticities  of our estimator and those of  \shortciteN{AMN:19} and \citeN{AW:22} based on a DGP with a trans-log production function. The last column contains the actual coverage probability of 95\% confidence interval for our estimator based on the bootstrap procedure in Section \ref{s:bootstrap}.
		\end{singlespace}} 
	\end{minipage}
	\end{center}
\end{table}  
\begin{table}[h!]
	\caption{Simulation results for quantile effects}
	\label{tab:sim_quantile_new_means_translog}
	\begin{center}
		\vspace{-1em}
		\begin{tabular}{l c ccc cc c}
			\hline \hline
			&  Bias  AMN  &  Std AMN  & Bias AW  & Std  AW & Bias AF & Std AF & Cov AF \\	
            \hline
			&\multicolumn{7}{c}{ $n = 500$} \\
			Skill effect  $t = 0$  & 0.0046 & 0.1210  & 0.0075 & 0.1364 & 0.0076 & 0.1162 & 0.920\\
			Skill effect  $t = 1$  & 0.0109 & 0.1330 & 0.0161 & 0.1460 & 0.0127 & 0.1339  & 0.878 \\
			Inv. effect  $t = 0$  & 0.0155 & 0.1355   & 0.0163 & 0.1465 & 0.0178 & 0.1313 & 0.917\\ 
			Inv. effect  $t = 1$  &  0.0342 & 0.1616 & 0.0342 & 0.1625 & 0.0354 & 0.1573  & 0.894 \\
			\hline
			& \multicolumn{7}{c}{ $n = 2000$} \\
			Skill effect  $t = 0$  & 0.0078 & 0.0630 & 0.0042 & 0.0718 & 0.0071 & 0.0582  & 0.932\\
			Skill effect  $t = 1$  & 0.0085 & 0.0679  & 0.0055 & 0.0778 & 0.0099 & 0.0672 & 0.921\\
			Inv. effect  $t = 0$  & 0.0072 & 0.0699  & 0.0069 & 0.0771 & 0.0064 & 0.0633 & 0.936\\ 
			Inv. effect  $t = 1$  &	 0.0065 & 0.0725   & 0.0088 & 0.0794 & 0.0057 & 0.0689 & 0.936 \\
			\hline
			\hline
		\end{tabular}
        \begin{minipage}{0.95\textwidth}
		\vspace{5mm}
		{\footnotesize \begin{singlespace} Notes:  The table shows biases and standard deviations of average quantile effects  of our estimator and those of  \shortciteN{AMN:19} and \citeN{AW:22} based on a DGP with a trans-log production function. The last column contains the actual coverage probability of 95\% confidence interval for our estimator based on the bootstrap procedure in Section \ref{s:bootstrap}.
		\end{singlespace}} 
	\end{minipage}
	\end{center}
\end{table}

\section{Empirical application}
\label{s:application}
The empirical application is based on \citeN{AW:22}. We first provide a brief outline of the model, which is slightly more complicated than the one in the previous section. The initial conditions are defined by the child's initial skills ($\theta_0$), the mother’s cognitive skills ($\theta_{MC}$), non-cognitive skills ($\theta_{MN}$), and initial family income ($Y_0$). Both cognitive and non-cognitive skills of the mother are assumed to be time-invariant. The distribution of these initial conditions is modeled as
\begin{align*}
    & (\ln \theta_0, \ln \theta_{MC}, \ln \theta_{MN}, \ln Y_0)   \sim \text{N}(\mu,  \Sigma).
\end{align*}
The evolution of skills follows a production function $f$, where current skills depend on past skills and parental investment
\begin{align*}
   & \ln \theta_{t+1} = f(\ln \theta_{t},\ln I_{t},\delta_{t}) + \eta_{\theta,t}.
\end{align*}
Parental investment is determined by current child skills, family income, and the mother’s cognitive and non-cognitive skills
\begin{align*}
    &\ln I_t = \beta_{0t} + \beta_{1t}\ln\theta_{t} + \beta_{2t}\ln \theta_{MN}+ \beta_{3t}\ln \theta_{MC} + \beta_{4t}\ln Y_t + \eta_{I,t}.
\end{align*}
Log family income ($Y_t$) follows an AR(1) process: $\ln Y_{t+1} = \nu_{Y,0} + \nu_{Y,1} \ln Y_{t} + \eta_{Y}$ for $t = 0,\ldots,T-1$.
Finally, the measurement system for the latent variables is given by
\begin{align*}
Z_{\theta,t,m} &= \mu_{\theta,t,m} + \lambda_{\theta,t,m} \ln \theta_t + \epsilon_{\theta,t,m} \hspace{26mm} \text{for all } t, m \\
Z_{I,t,m} &= \mu_{I,t,m} + \lambda_{I,t,m} \ln I_t + \epsilon_{I,t,m} \hspace{26mm} \text{for all } t, m \\
Z_{MC,m} &= \mu_{MC,m} + \lambda_{MC,m} \ln \theta_{MC} + \epsilon_{MC,m} \hspace{17mm} \text{for all } m \\
Z_{MN,m} &= \mu_{MN,m} + \lambda_{MN,m} \ln \theta_{MN} + \epsilon_{MN,m} \hspace{16mm} \text{for all } m.
\end{align*}
We use the same set of measures for children’s skills and the mother’s cognitive skills as \citeN{AW:22}. For children’s skills, we use the three scores from the Peabody Individual Achievement Test (PIAT) in Mathematics, Reading, and Recognition. For the mother’s cognitive skills, we rely on six measures derived from the Armed Services Vocational Aptitude Battery (ASVAB).

For the mother’s non-cognitive skills, we construct three "continuized" measures based on the 13 measures used in \citeN{AW:22}: first the average across the four Rotter indices, second the average of the five positively ordered Rosenberg indices (where a higher score indicates higher skills) and third the average of the four negatively ordered Rosenberg indices.  
All measures are coded such that higher scores tend to correspond to higher skills. Additional details can be found in Table B-7 of the web appendix of \citeN{AW:22}. For parental investment, we follow \citeN{AW:22} and use  three measures, namely “how often the mother reads to the child,” “how often the child is praised,” and “how often the child was taken to a museum.”

Given the significant number of missing values in the dataset, we use a complete subsample of the data and focus on ages 7, 9, and 11. Hence, $T = 2$. The resulting subsample consists of $1,403$ children.\footnote{\label{f:AW}As explained in their Appendix A.4, for their main results \citeN{AW:22} impute missing values. To do so, in the first stage of their two-stage least squares estimator, they regress each endogenous variable on a distinct subset of the instruments (i.e. the natural instrument for that variable) instead of all instruments. For instance, in the context of the production function, a skill measure is regressed on the remaining skill measures and an investment measure on the remaining investment measures. Even in the absence of missing data, such a procedure leads to inconsistent estimators because all measures are correlated. Using a complete subsample allows us to implement the estimator described in their main text. The authors will post a revised web appendix that also contains standard 2SLS results without imputations. Estimating such models with missing data can be challenging due to the large number of combinations of variables that have missing values, resulting in different parameters being estimated on different subsets of the data.}

We estimate the model using both a CES and a trans-log production function. For the CES specification, we present results for both our estimator and that of \shortciteN{AMN:19}. For the trans-log specification, we additionally include results for the IV estimator of \citeN{AW:22}. For our estimator and that of \shortciteN{AMN:19}, we use Assumption \ref{a:normalizations_tl} for the trans-log to achieve point identification of the primitive parameters. For the estimator of \shortciteN{AMN:19}, we impose Assumption \ref{a:normalizations_ces} for the CES case. For the CES estimator, our estimator assumes that $\beta_{0t} = 0$ instead of $\mu_{I,t,1} = 0$ for all $t$. \citeN{AW:22} impose a different set of identifying assumptions. For example, they set $\mu = (0,0,0, \mu_Y)$ and do not restrict the location parameters $\mu_{\theta, 0, 1}, \mu_{MN, 1}$ and $\mu_{MC, 1}$. Although the set of imposed assumptions does impact the parameter estimates, it does not have an effect on the reported features in this section \cite{Freyberger:24}. 

As outlined in Section \ref{s:aw}, \citeN{AW:22} make use of an IV strategy. In principle, each measure can be used in the main regression to either replace a latent variable or to serve as an instrument. We closely follow the choices of the main regressors and instruments of the implementation in \citeN{AW:22}. That is, in the investment equation, we use the PIAT math score as a proxy for latent skills. For cognitive and non-cognitive skills, we use "asvab2" and the average of the four negatively ordered Rosenberg indices, respectively. The set of instruments includes the other measures in the same period. For investment, we use “how often the child was taken to a museum” and “how often the child is praised”. For each of the two measures, we estimate the investment function parameters and then average the results, following \citeN{AW:22}. Since investment appears on the left-hand side of the equation, no instrument is required. 

For the production function, we follow \citeN{AW:22} and use the math scores as proxies for skills on both the left-hand and right-hand side. The measure used to replace investment differs by time period in \citeN{AW:22}. We follow the description of their estimator in the paper and use the same measure in each time period (as for the skills of the children). We report results for both “how often the child was taken to a museum” (denoted by ``AW museum'') and “how often the child is praised” (denoted by ``AW praised'') in the role of the investment-regressor in the production function. As instruments, we use the remaining measures in the same time period.

As explained in Footnote \ref{f:AW}, the implementation of \citeN{AW:22} differs from the description of the estimator in their paper, with the implemented estimator being generally inconsistent. The bias is visible in their Monte Carlo simulation, which uses the same estimation strategy without missing data. In addition to the estimator described in their paper (and in Section \ref{s:aw}), we also report results based on their actual implementation (denoted by ``AW 2025'') using “how often the child is praised” to replace investment in the production function.

\begin{figure}[b!]
	\begin{center}
	\caption{Median quantile effects: Trans-log production function}
	\label{f:quantiles_median_inv_application_trans}
	\includegraphics[trim=3cm 4.5cm 3cm 0cm, width=15cm]{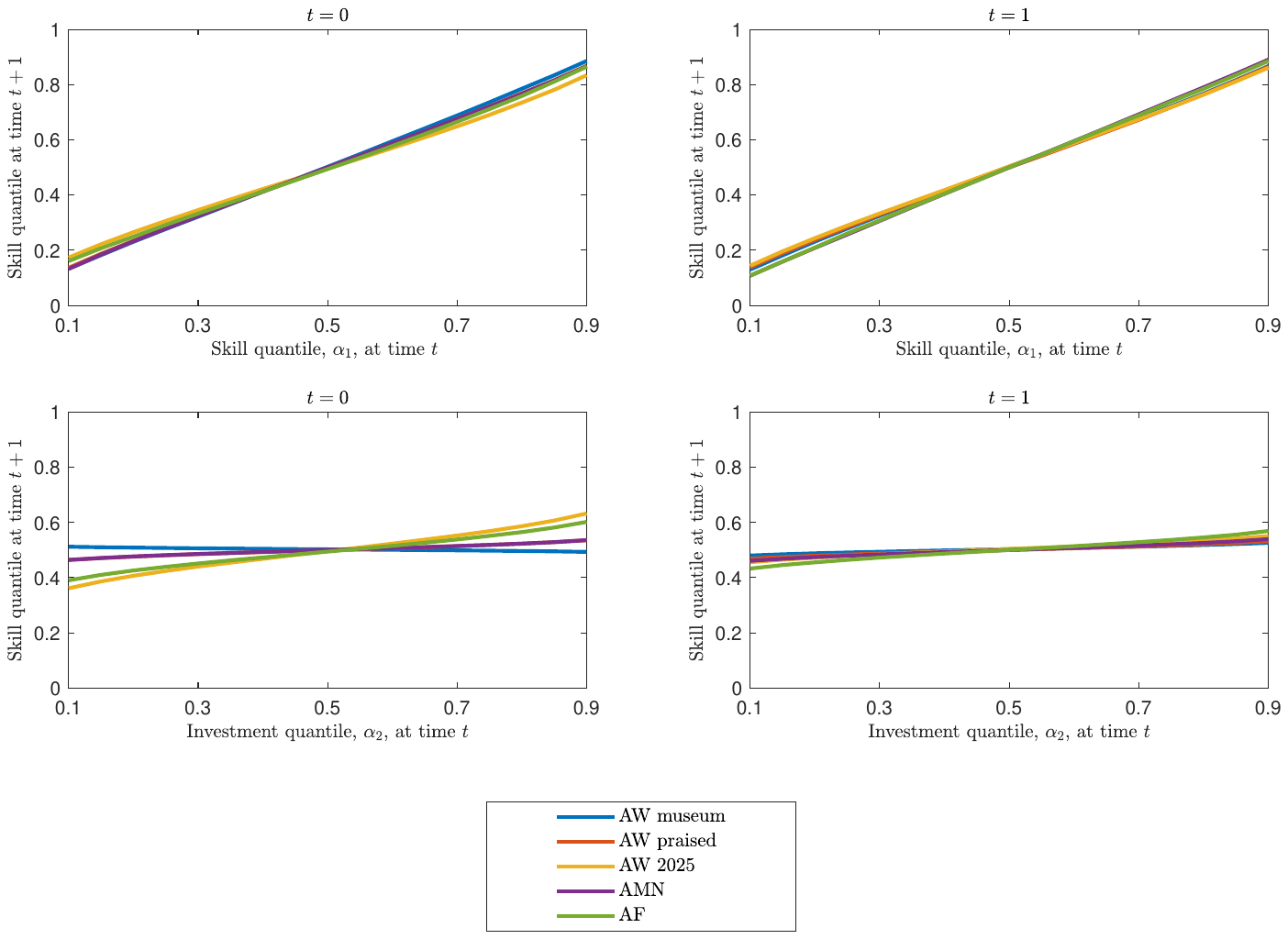}
    	\begin{minipage}{0.95\textwidth}
		\vspace{10mm}
		{\footnotesize \begin{singlespace} Notes: The figure shows estimated average quantile effects in the trans-log specification of our estimator, that of \shortciteN{AMN:19} and different implementations of \citeN{AW:22}. For the upper panels, $\alpha_2 = 0.5$ and for the lower panels $\alpha_1 = 0.5$. 
		\end{singlespace}} 
	\end{minipage}
		\end{center}
\end{figure}

We now report results on identified features similar to those reported in the Monte Carlo study. Figure \ref{f:quantiles_median_inv_application_trans} shows estimates of 
\begin{align*}
F_{\theta_{t+1}}\left(a_t + \gamma_{1t}\ln Q_{\alpha_1}(\theta_{t}) + \gamma_{2t} \ln Q_{\alpha_2}(I_{t}) + \gamma_{3t}\ln Q_{\alpha_1} (\theta_{t}) \ln Q_{\alpha_2} (I_{t}) \right).
\end{align*}
In the upper panels, investment is fixed to the median level, i.e. $\alpha_2 = 0.5$, and the quantile of skills, $\alpha_1$, varies. For the lower panels, $\alpha_1 = 0.5$ and we show results for different values of $\alpha_2$. We set all unobservables to their median values. Most estimates yield similar quantile effects. However, the dynamics in the lower left panel depend on the exact specification used for the estimator of \citeN{AW:22}. Specifically, \citeN{AW:22} implemented with “how often the child was taken to a museum” (blue line) indicates a weak negative relationship between investment and skills in the next period, whereas all other curves indicate a (weak) positive relationship. This relationship is less pronounced for the estimator of \shortciteN{AMN:19}.

\begin{figure}[b!]
	\begin{center}
	\caption{Counterfactual distributions: Trans-log production function}
	\label{f:distributions_investment_application_trans}
	\includegraphics[trim=2cm 0.5cm 3cm 0cm, width=15cm]{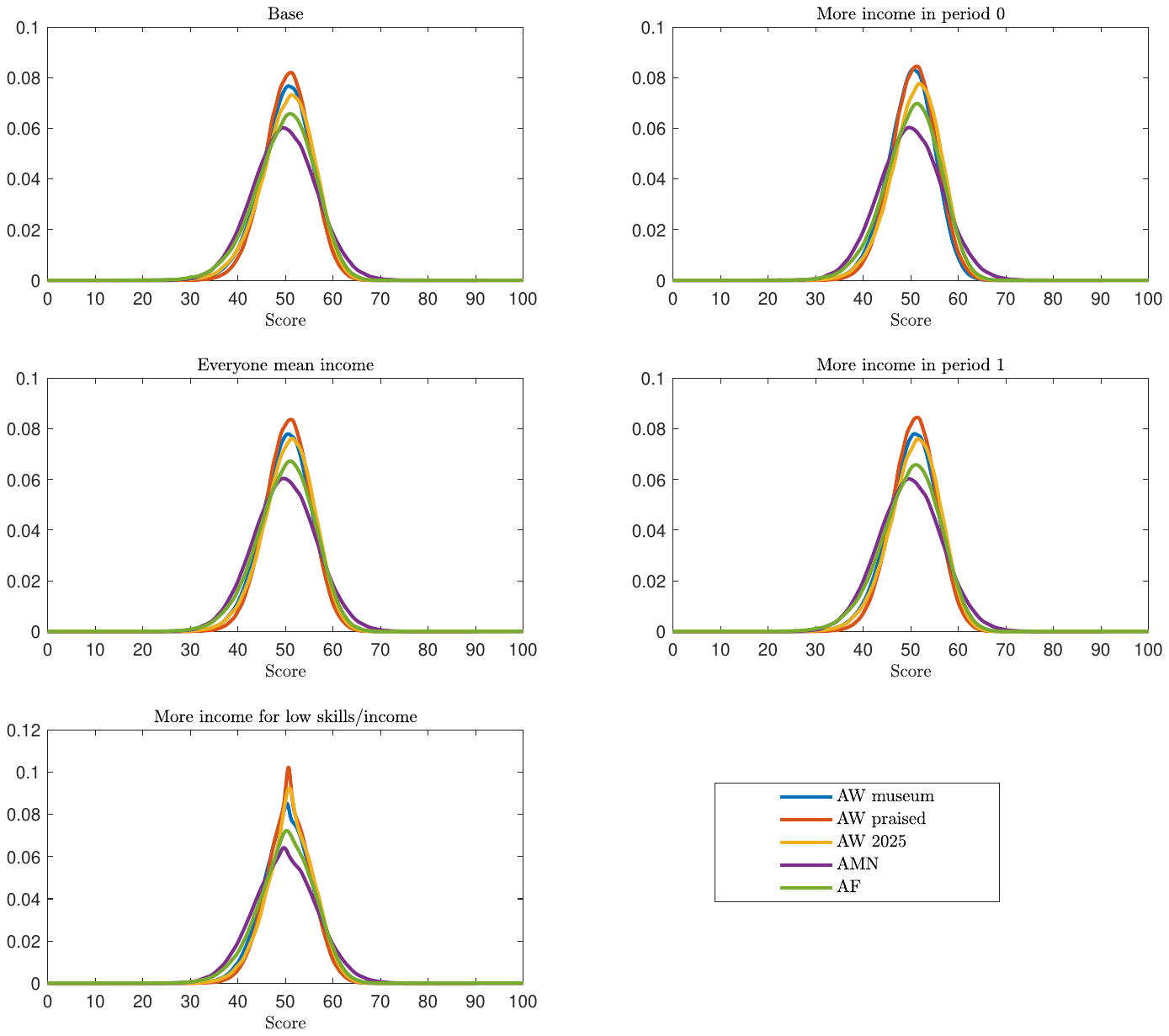}
    	\begin{minipage}{0.85\textwidth}
		\vspace{5mm}
		{\footnotesize \begin{singlespace} Notes: The figure shows the estimated average counterfactual distributions of the PIAT math measure for $t = 2$.
		\end{singlespace}} 
	\end{minipage}
		\end{center}
\end{figure}

Figure \ref{f:distributions_investment_application_trans} shows averages of the estimated counterfactual math score distributions for the final period under different counterfactual income distributions. As in the simulations, we (1) increase everyone's income by two standard deviations in period 0, (2) increase everyone's income by two standard deviations in period 1, (3) set income to the median for everyone in both periods, and (4) increase income by two standard deviations in both periods, but only if the initial skill and income quantiles are below $0.5$. Again, we set all unobservables to their median values. All estimators yield quite similar counterfactual distributions.

\begin{figure}[t!]
	\begin{center}
	\caption{Quantile paths: Trans-log production function}
	\label{f:quantile_paths_application_trans}
	\includegraphics[trim=4cm 8.5cm 4cm 4cm, width=15cm]{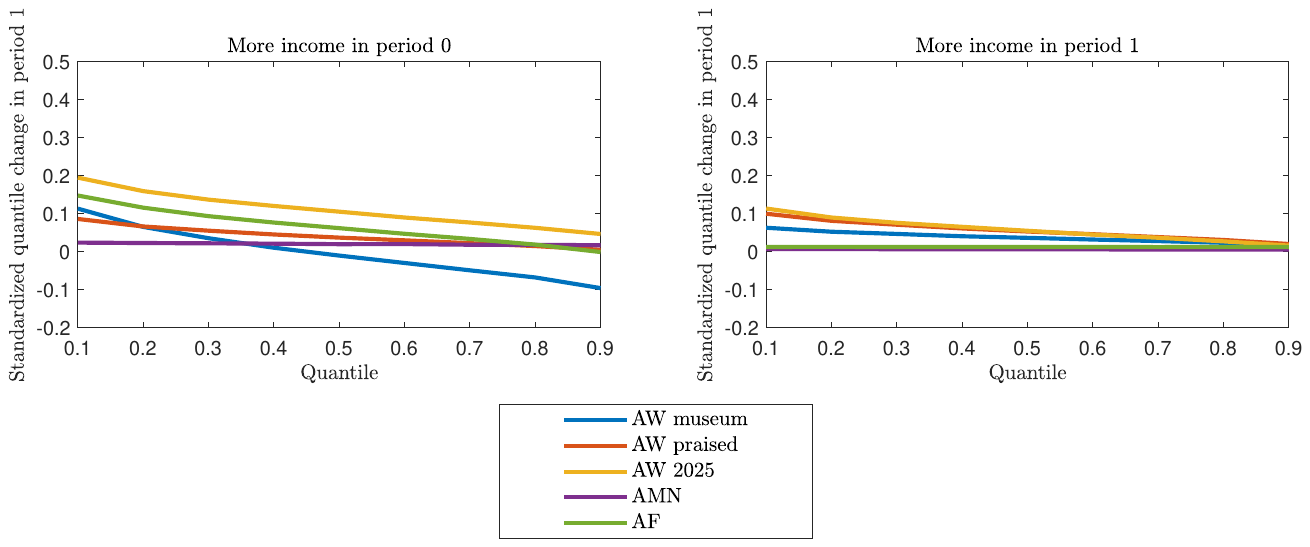}
    	\begin{minipage}{0.95\textwidth}
		\vspace{5mm}
		{\footnotesize \begin{singlespace} Notes:  The figure shows the standardized change in the quantile of the skill distributions
        for transfers in $t = 0$ and $t = 1$ based on the trans-log production function.
		\end{singlespace}} 
	\end{minipage}
		\end{center}
\end{figure}

Just like in the simulations, we now consider the timing of investment - either in period 0 or in period 1 - in more detail. Figure \ref{f:quantile_paths_application_trans}  shows the standardized changes in the quantiles of the skill distributions for earlier and later transfers. Typically, the estimated effects of income on skills are non-negative for all quantiles. An exception is the estimates of \citeN{AW:22} with “how often the child was taken to a museum”, which show a positive effect for low quantiles and a negative effect for high quantiles in the left panel. The conclusion for the optimal timing of investment might therefore depend on the exact specification.  Our estimator yields a small positive  effect in period $0$ and a negligible positive effect in period $1$. The estimates of \shortciteN{AMN:19} imply a negligible positive effect in period $0$ and no effect in period $1$.  

\begin{figure}[h]
	\begin{center}
	\caption{Median quantile effects: CES production function}
	\label{f:quantiles_median_inv_application_ces}
	\includegraphics[trim=3cm 4.0cm 3cm 0cm, width=15cm]{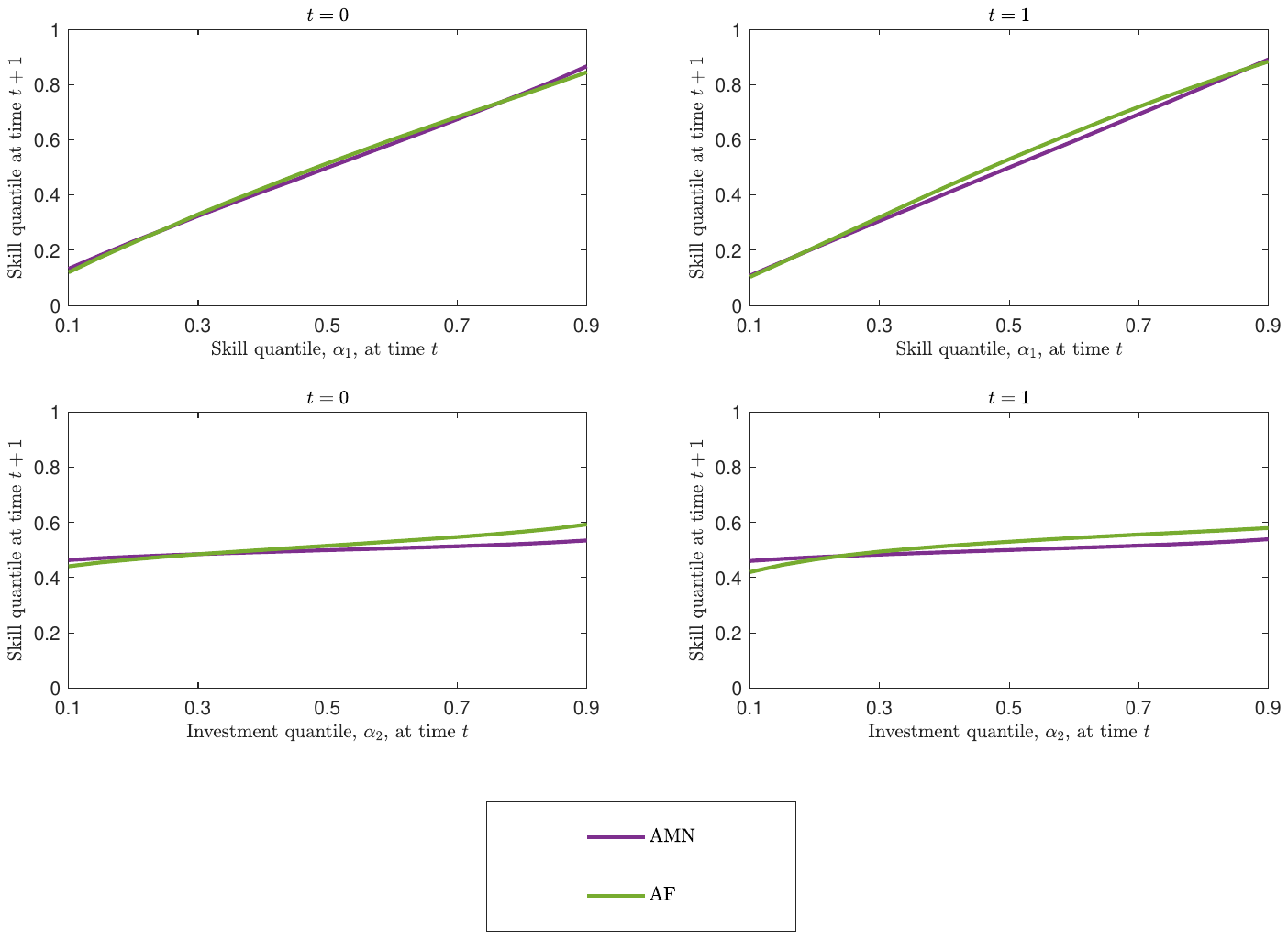}
    	\begin{minipage}{0.95\textwidth}
		\vspace{5mm}
		{\footnotesize \begin{singlespace} Notes: As Figure \ref{f:quantiles_median_inv_application_trans}, but for the CES production function. Here, we report results for our estimator and that of \shortciteN{AMN:19}.
		\end{singlespace}} 
	\end{minipage}
		\end{center}
\end{figure}

The results for the CES specification are similar. Figure \ref{f:quantiles_median_inv_application_ces} displays the counterpart of Figure \ref{f:quantiles_median_inv_application_trans}. For the corresponding counterparts of Figures \ref{f:distributions_investment_application_trans} and \ref{f:quantile_paths_application_trans} see Figures \ref{f:distributions_investment_application_ces} and \ref{f:quantile_paths_application_ces}  in Appendix \ref{s:appendix_fig}. 
Compared to Figure \ref{f:quantiles_median_inv_application_trans} for the trans-log production function, Figure \ref{f:quantiles_median_inv_application_ces} indicates a weaker positive relationship between current parental investment and future skills for our estimator. However, we should note that the estimated share parameter of investment ($\gamma_{2t}$ in Equation \ref{eq:ces}) is close to zero for $t = 1$.\footnote{The primitive parameter $\gamma_{2t}$ is not identified \cite{Freyberger:24}. Hence, interpreting it is not possible and potentially misleading. However, Figure \ref{f:quantiles_median_inv_application_ces} also indicates that skills are quite persistent and that investment has only little impact on future skills.} The CES production function is only weakly identified if one of the share parameters ($\gamma_{1t}$ and $\gamma_{2t}$ in Equation \ref{eq:ces}) is close to $0$. To see this, consider the CES production function and assume that $\gamma_{2t} = 0$, then 
$$\theta_{t+1} =  ( \gamma_{1t} \theta_{t}^{\sigma_{t}} + \gamma_{2t} I_{t}^{\sigma_{t}} )^{\psi_t/\sigma_{t}} \exp(\eta_{\theta,t}) =  \gamma_{1t}^{1/\sigma_{t}} (\theta_t )^{\psi_t} \exp(\eta_{\theta,t}).$$
It follows that $\sigma_t$ and $\gamma_{1t}$ are not separately identified. Weak identification may lead to computational instability, which has been the case here - see the following section for further remarks.

\section{Discussion and practical recommendations}
\label{s:discussion}

In this section, we provide additional practical recommendations. 

\textbf{Starting values:} Our estimation procedure, like any other optimization-based method, can be sensitive to the choice of starting values. In principle, researchers should explore different starting values to ensure robust results. While our procedure remains computationally feasible, it does require numerical integration, which can be computationally intensive. As a result, testing numerous starting values without clear guidance may be impractical. In this setting, the estimates from \shortciteN{AMN:19} are a natural candidate for starting values. Although their estimator relies on assumptions that may not align with the model, it is easy to implement, it is fast, and it can provide a reasonable approximation to guide the optimization process. Another approach for obtaining starting values is to use our estimator but with numerical integration performed on a reduced number of nodes, as they often result in reasonable starting values.

\textbf{Weak identification:} With the CES production function, all estimators can be numerically instable and can have poor statistical properties when the parameters are only weakly identified in the sense that the true parameter vector is close to a vector for which point identification fails. There are at least three distinct sources of weak identification. First, identification of the CES production function under Assumptions \ref{a:baseline} and \ref{a:normalizations_ces} requires that $\gamma_{1t}, \gamma_{2t}$ and $\sigma_t$ are unequal to $0$ for all $t$ \cite{Freyberger:24}. If one of the parameters is close to zero, the true parameter vector is only weakly identified. Identification failure, when $\gamma_{1t} = 0$ or $\gamma_{2t}= 0$, is mentioned at the end of Section \ref{s:application}. Second, if $\sigma_t$ is close to $0$, the CES approaches the Cobb-Douglas production function. As the Cobb-Douglas is a special case of the trans-log production function, identification requires additional scaling restrictions (see Assumption \ref{a:normalizations_tl} vs. \ref{a:normalizations_ces}). Third, if the number of mixtures is over-specified, identification fails and estimators are unstable, as discussed in Section \ref{s:simulations}.  

\textbf{Functional form:} Recall that the Cobb-Douglas production function is the limit of the CES production function as $\sigma_t \rightarrow 0$. When the trans-log production also includes $(\ln \theta_t)^2$ and $(\ln I_t)^2$ as additional terms, it is a first order approximation of the CES production function around $\sigma_t = 0$. Even though the CES and the trans-log production function are nonnested, the trans-log production function appears more restrictive. For example, for the trans-log production function without quadratic terms, $\frac{ \partial \ln \theta_{t+1}}{\partial \ln \theta_t}$ does not depend on the level of skills in period $t$ and thus, all the lines in Figure \ref{f:elasticities_original} would be horizontal. However, one advantage of the trans-log production  function is that it avoids the weak identification issues discussed above. One way to achieve  both sufficient flexibility and numerical stability could be to add higher order terms to the trans-log production function.

\textbf{Standardizations:} For the CES specification, the performance of the optimizer also depends on the scale and the location of the measures. To see this, suppose for simplicity that investment is observed, and $\ln I_t = Z_{I,t}$. The summary statistics and counterfactuals reported in the previous sections are invariant to changes in the units of measurement of the data.  However, scaling the data can affect the numerical performance of our method. To see why, consider the CES production function
\begin{align*}
    \theta_{t+1} &= \left( \gamma_{1t} \theta_{t}^{\sigma_{t}} + \gamma_{2t} I_{t}^{\sigma_{t}} \right)^{\psi_t/\sigma_{t}} \exp(\eta_{\theta,t}) \\
     &= \left( \gamma_{1t} \theta_{t}^{\sigma_{t}} + \gamma_{2t} \exp(Z_{I,t})^{\sigma_{t}} \right)^{\psi_t/\sigma_{t}} \exp(\eta_{\theta,t}).
\end{align*}
If $\sigma_t > 0$, the right hand side may be numerically unstable if $Z_{I,t}$ contains large values. Standardizing the measures is a practical way to avoid these issues and improve the stability of the optimization process.

\textbf{Binary measures:} A major advantage of the likelihood-based approach is that it naturally allows for binary measures, which are common in applications. In this case, one could for example assume that a binary skill measure $Z_{\theta,t,m}$ can be written as
$$Z_{\theta,t,m} = \1(\mu_{\theta,t,m} + \lambda_{\theta,t,m} \ln \theta_t \geq \eps_{\theta,t,m})$$
instead of 
$$Z_{\theta,t,m} = \mu_{\theta,t,m} + \lambda_{\theta,t,m} \ln \theta_t + \eps_{\theta,t,m}.$$
Assuming a probit model with $\eps_{\theta,t,m} \mid  \ln \theta_t \sim N(0, 1)$, we then get
$$P(Z_{\theta,t,m} = 1  \mid  \ln \theta_t) = \Phi(\mu_{\theta,t,m} + \lambda_{\theta,t,m} \ln \theta_t)$$
which can be incorporated in the likelihood, analogous to $f_{Z_{\theta,t,m} \mid \ln \theta_t}$ in the continuous case. Contrarily, assuming that these measures are distributed as mixtures of normals might yield particularly poor approximations. \shortciteN{CHS:10} allow for discrete measures of adult outcomes in their identification. However, it remains unclear how other discrete measures can be incorporated into their estimation framework, given that the underlying approximation continues to rely on a normal distribution or a mixture of normals.

 \textbf{Numerical integration:} As mentioned in Section \ref{s:likelihood}, we use numerical integration to evaluate the likelihood, and we experimented with different methods. Among these, we found that quasi Monte Carlo integration based on Halton sequences performed particularly well in the Monte Carlo simulations. In the simulations, we use $10,000$ draws to evaluate each integral. In the application, using either $10,000$ or $20,000$ draws yields essentially identical results.

\textbf{Missing data:}  Missing values in the measures are conceptually straightforward to incorporate in our likelihood under the assumption of missing at random. In such cases, for each observation, we can construct the contribution to the likelihood function based only on the observed data, which is then a function of a subset of the full parameter vector. Depending on how many different combinations of measures are missing, the primary challenge lies in implementing each of these individual contributions.

\bibliography{references}

\clearpage 

\appendix

\setcounter{figure}{0}
\setcounter{table}{0}
\setcounter{theorem}{0}
\setcounter{assumption}{0}
\renewcommand{\thetable}{A.\arabic{table}}
\renewcommand{\thefigure}{A.\arabic{figure}}
\renewcommand{\thetheorem}{A.\arabic{theorem}}
\renewcommand{\theassumption}{A.\arabic{assumption}}
	
	\allowdisplaybreaks

\section{Additional tables and figures}
\label{s:appendix_fig}

\begin{figure}[h]
	\begin{center}
	\caption{Counterfactual distributions: CES production function}
	\label{f:distributions_investment_application_ces}
	\includegraphics[trim=2cm 0.5cm 3cm 0cm, width=15cm]{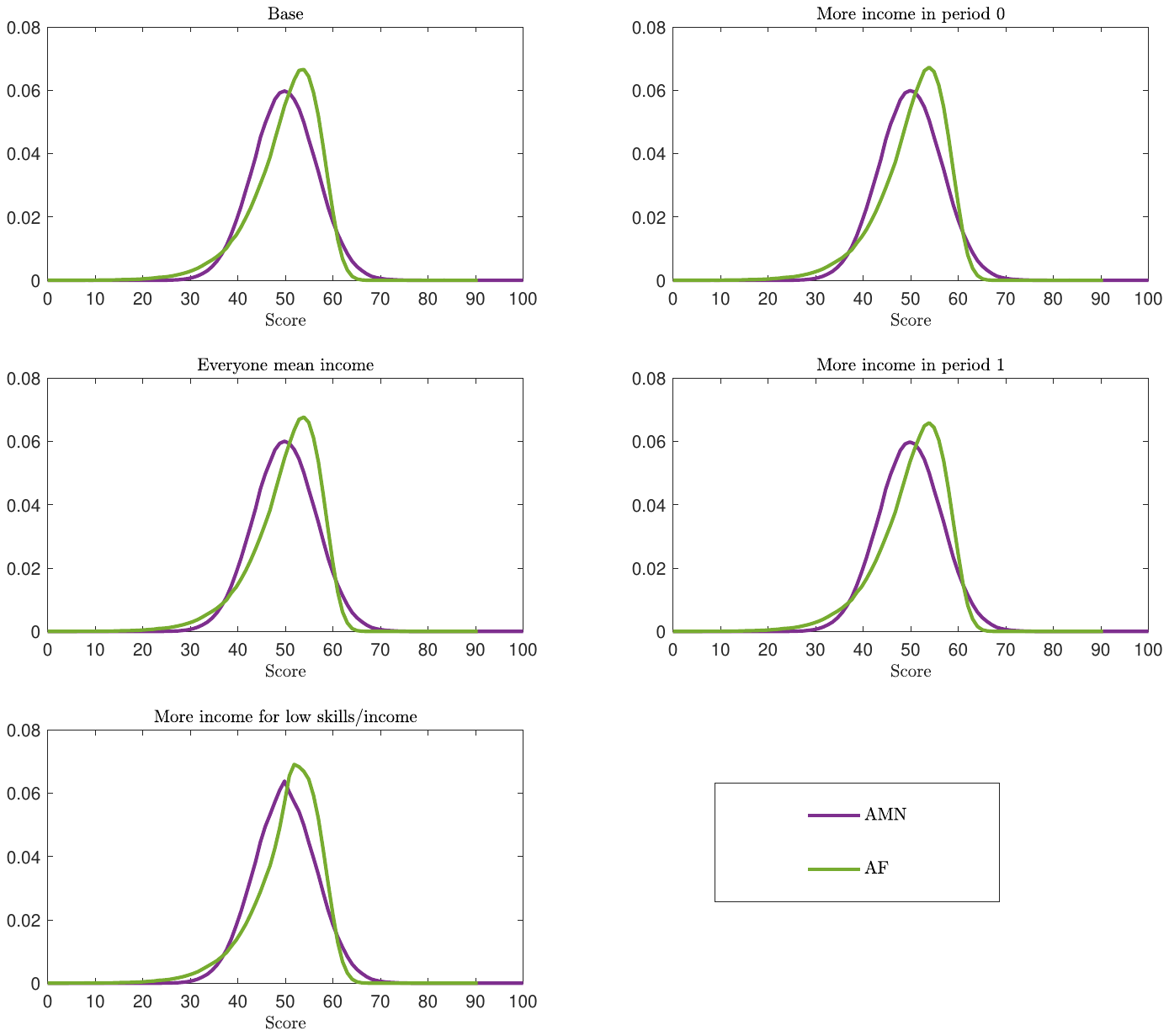}
    	\begin{minipage}{0.8\textwidth}
		\vspace{5mm}
		{\footnotesize \begin{singlespace} Notes: As Figure \ref{f:distributions_investment_application_trans}, but for the CES production function.
		\end{singlespace}} 
	\end{minipage}
		\end{center}
\end{figure}

\begin{figure}[h]
	\begin{center}
	\caption{Quantile paths: CES production function}
	\label{f:quantile_paths_application_ces}
	\includegraphics[trim=4cm 8.5cm 4cm 4cm, width=15cm]{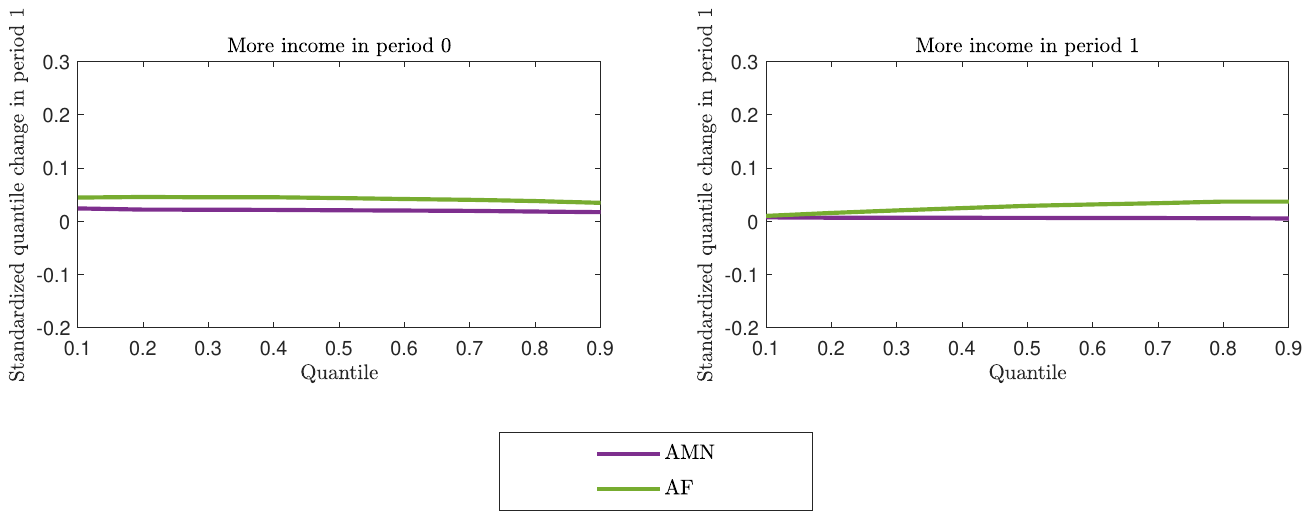}
    	\begin{minipage}{0.95\textwidth}
		\vspace{5mm}
		{\footnotesize \begin{singlespace} Notes: As Figure \ref{f:quantile_paths_application_trans}, but for the CES production function.
		\end{singlespace}} 
	\end{minipage}
		\end{center}
\end{figure}

\end{document}